\documentclass[runningheads]{llncs}
\usepackage{paralist}
\usepackage{units}
\usepackage{booktabs} 
\usepackage{graphicx}
\usepackage{latexsym}
\usepackage{multirow}
\usepackage{booktabs}
\usepackage{xspace}
\usepackage{caption}
\usepackage{float}
\usepackage{wrapfig}
\usepackage{subcaption}

\usepackage[breaklinks]{hyperref}
\hypersetup{
colorlinks = true, 
linkcolor=blue, 
citecolor=blue, 
urlcolor=blue 
}

\usepackage{algorithm}
\usepackage[noend]{algpseudocode} 
\algnewcommand\algorithmicswitch{\textbf{switch}}
\algnewcommand\algorithmiccase{\textbf{case}}
\algdef{SE}[SWITCH]{Switch}{EndSwitch}[1]{\algorithmicswitch\ #1\ \algorithmicdo}{\algorithmicend\ \algorithmicswitch}%
\algdef{SE}[CASE]{Case}{EndCase}[1]{\algorithmiccase\ #1}{\algorithmicend\ \algorithmiccase}%
\algtext*{EndSwitch}%
\algtext*{EndCase}%
\algdef{SE}{PreLoop}{EndPreLoop}[1]{\textbf{preloop} \(\mbox{#1}\) \textbf{do}}{\textbf{end}}%
\newcommand{\Cstate}[2]{\State \textbf{case} #1 #2}
\algnewcommand{\IIf}[1]{\State\algorithmicif\ #1\ \algorithmicthen}
\algnewcommand{\EndIIf}{\unskip\ 
}

\usepackage{amsmath}

\usepackage{amssymb}

\usepackage{graphicx}
\usepackage{color}
\usepackage{colortbl}
\usepackage{tcolorbox}
\usepackage{mathrsfs}

\usepackage{pgf}
\usepackage{tikz}
\usetikzlibrary{arrows.meta,matrix,shapes,decorations,automata,backgrounds,positioning}

\newcommand\mathcompact
{%
\thinmuskip= 0mu                  
\medmuskip= 0mu 
\thickmuskip= 0mu        
}

\newcommand{\true}{\ensuremath{\mathtt{true}}\xspace}
\newcommand{\false}{\ensuremath{\mathtt{false}}\xspace}
\newcommand{\R}{\mathbb{R}}
\newcommand{\N}{\mathbb{N}}
\newcommand{\Rpos}{\R_{+}}
\newcommand{\Npos}{\N_{+}}

\newcommand{\bw}{\ensuremath{\mathbf{v}}\xspace}

\newcommand{\bwp}[2]{\ensuremath{\mathbf{v}_{#1:#2}}\xspace}

\newcommand{\Var}{\mathbf{Var}}

\newcommand{\UntilOp}[1]{\mathbin{\mathcal{U}_{#1}}}

\newcommand{\Rnn}{\R_{\ge 0}}

\newcommand{\Defeq}{:=}

\newcommand{\Robust}[3]{\mathrm{R}(#1, #2, #3)}

\newcommand{\Rmax}{\mathtt{R}^{\alpha}_{\mathrm{max}}}
\newcommand{\Rmin}{\mathtt{R}^{\alpha}_{\mathrm{min}}}
\newcommand{\Mon}[3]{\mathrm{[R]}(#1, #2, #3)}
\newcommand{\MonL}[3]{\mathrm{[R]}^\mathsf{L}(#1, #2, #3)}
\newcommand{\MonU}[3]{\mathrm{[R]}^\mathsf{U}(#1, #2, #3)}
\newcommand{\MonUnoArgs}{\ensuremath{\mathrm{[R]}^\mathsf{U}}\xspace}
\newcommand{\MonLnoArgs}{\ensuremath{\mathrm{[R]}^\mathsf{L}}\xspace}
\newcommand{\InsRobSat}[3]{\mathcal{[\mathscr{R}]}^\oplus\left(#1, #2, #3\right)}
\newcommand{\InsRobVio}[3]{\mathcal{[\mathscr{R}]}^\ominus\left(#1, #2, #3\right)}
 \newcommand{\DiagTSimp}[3]{\mathrm{E^\oplus}(#1, #2, #3)}
 \newcommand{\CDiag}[3]{\mathscr{M}(#1, #2, #3)}
 \newcommand{\DiagFSimp}[3]{\mathrm{E^\ominus}(#1, #2, #3)}
\newcommand{\CMon}[3]{\mathrm{M}(#1, #2, #3)}

\newcommand{\alphaset}{\mathcal{A}}

\newcommand{\stepLength}{\ensuremath\delta\xspace}

\newcommand{\unknown}{\ensuremath{\mathord{?}}\xspace}
\newcommand{\textunknown}{\ensuremath{\mathtt{unknown}}\xspace}
\newcommand{\posCause}{\ensuremath{\oplus}\xspace}
\newcommand{\negCause}{\ensuremath{\ominus}\xspace}
\newcommand{\nothing}{\ensuremath{\oslash}\xspace}
\newcommand{\speed}{\ensuremath{\mathtt{speed}}\xspace}
\newcommand{\gear}{\ensuremath{\mathtt{gear}}\xspace}
\newcommand{\rpm}{\ensuremath{\mathtt{RPM}}\xspace}
\newcommand{\acceleration}{\ensuremath{\mathtt{acceleration}}\xspace}

\newcommand{\af}{\ensuremath{\mathtt{AF}}\xspace}
\newcommand{\afref}{\ensuremath{\mathtt{AFref}}\xspace}

\newcommand{\timeSymbol}{\ensuremath{b}\xspace}

\newcommand{\spec}[2]{\ensuremath{\varphi^{\mathsf{#1}}_{#2}}\xspace}
\newcommand{\sig}[1]{\ensuremath{\#{#1}}\xspace}

\newcommand{\trans}[2]{\ensuremath{\Gamma(#1, #2)}\xspace}

\newcommand{\cnf}{\mathrm{CNF}}
\newcommand{\dnf}{\mathrm{DNF}}
\newcommand{\cnfclause}{c}
\newcommand{\dnfclause}{d}
\newcommand{\cnfClauseSet}{C}
\newcommand{\dnfClauseSet}{D}
\newcommand{\dimension}{\mathit{d}}

\newcommand{\cto}{\sqsubset_{\mathsf{c}}}
\newcommand{\dto}{\sqsubset_{\mathsf{d}}}
\newcommand{\concTime}[1]{\ensuremath{\mathsf{ConT}(#1)}\xspace}
\newcommand{\lowRes}[1]{\mathfrak{l}_{#1}}
\newcommand{\uppRes}[1]{\mathfrak{u}_{#1}}

\newcommand{\classicMon}{\texttt{ClaM}\xspace}
\newcommand{\resetMon}{\texttt{ResM}\xspace}
\newcommand{\boolCauseMon}{\texttt{BCauM}\xspace}
\newcommand{\quanCauseMon}{\texttt{QCauM}\xspace}

\newcommand{\myparagraph}[1]{\medskip\noindent{\bf #1.}}
\allowdisplaybreaks

\spnewtheorem{mytheorem}{Theorem}
{\bfseries}{\rmfamily} 
\spnewtheorem{mylemma}{Lemma}
{\bfseries}{\rmfamily} 
\spnewtheorem{myexample}{Example}{\bfseries}{\rmfamily}
\spnewtheorem{myassumption}[mytheorem]{Assumption}{\bfseries}{\rmfamily}
\spnewtheorem{mydefinition}{Definition}{\bfseries}{\rmfamily}
\spnewtheorem{myremark}{Remark}{\bfseries}{\rmfamily}

\usepackage{cite}

\usepackage{extarrows}
\newcommand{\mycomment}[1]{\footnotesize{\text{(#1)}}}

\makeatletter
\def\orcidID#1{\kern .08em\href{https://orcid.org/#1}{\includegraphics[keepaspectratio,width=0.9em]{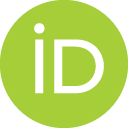}}}
\makeatother

\usepackage[firstpage]{draftwatermark}  

\SetWatermarkAngle{0}
\SetWatermarkText{\raisebox{12.5cm}{%
  \hspace{0.1cm}%
  \href{https://doi.org/10.5281/zenodo.7923888}{\includegraphics{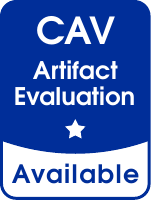}}%
  \hspace{9cm}%
  \includegraphics{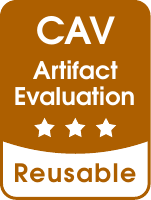}%
}}

\begin{document}

\title{Online Causation Monitoring of\\Signal Temporal Logic\thanks{Z. Zhang is supported by JSPS KAKENHI Grant No. 23K16865 and No. 23H03372. J. An, P. Arcaini, and I. Hasuo are supported by ERATO HASUO Metamathematics for Systems Design Project (No. JPMJER1603), JST, Funding Reference number 10.13039/501100009024 ERATO. P.Arcaini is also supported by Engineerable AI Techniques for Practical Applications of High-Quality Machine Learning-based Systems Project (Grant Number JPMJMI20B8), JST-Mirai.}}
\titlerunning{Online Causation Monitoring of STL}
%
\author{Zhenya Zhang\inst{1}\orcidID{0000-0002-3854-9846}
\and
Jie An\inst{2}\orcidID{0000-0001-9260-9697}
\and
Paolo Arcaini\inst{2}\orcidID{0000-0002-6253-4062}
\and
Ichiro Hasuo\inst{2}\orcidID{0000-0002-8300-4650}
}

\authorrunning{Z. Zhang, J. An, P. Arcaini, I. Hasuo}

\institute{Kyushu University, Fukuoka, Japan \\
\email{zhang@ait.kyushu-u.ac.jp}
\and 
National Institute of Informatics, Tokyo, Japan \\ 
\email{\{jiean,arcaini,hasuo\}@nii.ac.jp}
}

\maketitle 
\begin{abstract}
Online monitoring is an effective validation approach for hybrid systems, that, at runtime, checks whether the (partial) signals of a system satisfy a specification in, e.g., \emph{Signal Temporal Logic (STL)}. The classic STL monitoring is performed by computing a robustness interval that specifies, at each instant, how far the monitored signals are from violating and satisfying the specification. However, since a robustness interval monotonically shrinks during monitoring, classic online monitors may fail in reporting new violations or in precisely describing the system evolution at the current instant. In this paper, we tackle these issues by considering the \emph{causation} of violation or satisfaction, instead of directly using the robustness. We first introduce a \emph{Boolean causation monitor} that decides whether each instant is relevant to the violation or satisfaction of the specification. We then extend this monitor to a \emph{quantitative causation monitor} that tells how far an instant is from being relevant to the violation or satisfaction. We further show that classic monitors can be derived from our proposed ones. Experimental results show that the two proposed monitors are able to provide more detailed information about system evolution, without requiring a significantly higher monitoring cost.
\end{abstract}

\section{Introduction}\label{sec:intro}

Safety-critical systems require strong correctness guarantees. Due to the complexity of these systems, offline verification may not be able to guarantee their total correctness, as it is often very difficult to assess all possible system behaviors. To mitigate this issue, runtime verification~\cite{Leucker2009,SanchezSABBCFFK19,RVlectures} has been proposed as a complementary technique that analyzes the system execution at runtime. Online monitoring is such an approach  that checks whether the system execution (e.g., given in terms of signals) satisfies or violates a system specification specified in a temporal logic~\cite{pnueli1977temporal,Koymans90}, e.g., \emph{Signal Temporal Logic (STL)}~\cite{maler2004monitoring}.

\emph{Quantitative online monitoring} is based on the
STL \emph{robust semantics}~\cite{fainekos2009robustness,donze2010robust} that not only tells whether a signal satisfies or violates a specification $\varphi$ (i.e., the classic Boolean satisfaction relation), but also assigns a value in $\R \cup \{\infty,-\infty\}$ (i.e., \emph{robustness}) that indicates \emph{how robustly} $\varphi$ is satisfied or violated. However, differently from offline assessment of STL formulas, an online monitor needs to reason on \emph{partial signals} and, so, the assessment of the robustness should be adapted. We consider an established approach~\cite{deshmukh2017robust} employed by \emph{classic online monitors} (\classicMon in the following). It consists in computing, instead of a single robustness value, a \emph{robustness interval}; at each monitoring step, \classicMon identifies an \emph{upper bound} $\MonUnoArgs$ telling the maximal reachable robustness of any possible suffix signal (i.e., any continuation of the system evolution), and a \emph{lower bound} \MonLnoArgs telling the minimal reachable robustness. If, at some instant, \MonUnoArgs becomes negative, the specification is violated; if \MonLnoArgs becomes positive, the specification is satisfied. In the other cases, the specification validity is \textunknown.

\begin{wrapfigure}[13]{r}{0.45\textwidth}
\vspace{-15pt}
\includegraphics[width=\linewidth]{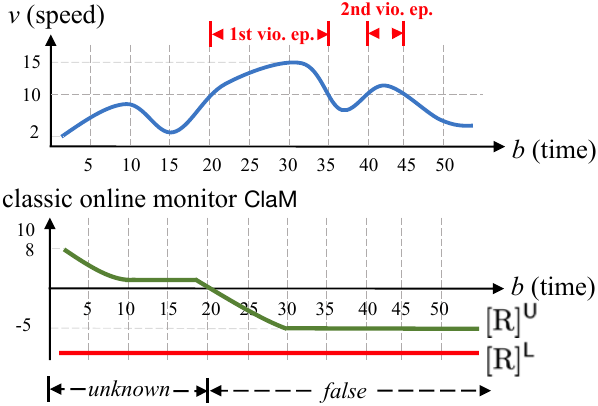}
\caption{\classicMon{} -- Robustness upper and lower bounds of of $\Box_{[0,100]}(v < 10)$}
\label{fig:runningExampleIntro}
\end{wrapfigure}
%
Consider a simple example in Fig.~\ref{fig:runningExampleIntro}. It shows the monitoring of the \speed of a vehicle (in the upper plot); the specification requires the \speed to be always below $10$. The lower plot reports how the upper bound \MonUnoArgs and the lower bound \MonLnoArgs of the reachable robustness change over time. We observe that the initial value of \MonUnoArgs is around 8 and gradually decreases.\footnote{The value of lower bound \MonLnoArgs is not shown in the figure, as not relevant. In the example, it remains constant before $b=100$, and the value is usually set either according to domain knowledge about system signals, or to $-\infty$ otherwise.} 
The monitor allows to detect that the specification is violated at time $\timeSymbol = 20$ when the \speed becomes higher than 10, and therefore \MonUnoArgs goes below 0. After that, the violation severity progressively gets worse till time $\timeSymbol = 30$, when \MonUnoArgs becomes $-5$. After that point, the monitor does not provide any additional useful information about the system evolution, as \MonUnoArgs remains stuck at $-5$. However, if we observe the signal of the \speed after $\timeSymbol = 30$, we notice that
\begin{inparaenum}[(i)]
\item the severity of the violation is mitigated, and the ``1st violation episode'' ends at time $\timeSymbol = 35$; however, the monitor \classicMon does not report this type of information;
\item a ``2nd violation episode'' occurs in the time interval $[40, 45]$; the monitor \classicMon does not distinguish the new violation.
\end{inparaenum}

The reason for the issues reported in the example is that the upper and lower bounds are monotonically decreasing and increasing; this has the consequence that the robustness interval at a given step is ``masked'' by the history of previous robustness intervals, and, e.g., it is not possible to detect mitigation of the violation severity. Moreover, as an extreme consequence, as soon as the monitor \classicMon assesses the violation of the specification (i.e., the upper bound \MonUnoArgs becomes negative), or its satisfaction (i.e., the lower bound \MonLnoArgs becomes positive), the Boolean status of the monitor does not change anymore. Such characteristic directly derives from the STL semantics and it is known as the \emph{monotonicity}~\cite{cimatti2019assumption,DeckerRV2013,ciccone2022ain}  of classic online monitors. Monotonicity has been recognized as a problem of these monitors in the literature~\cite{cimatti2019assumption,selyunin2017runtime,onlineResetTCAD2022}, since it does not allow to detect specific types of information that are ``masked''. We informally define two types of \emph{information masking} that can occur because of monotonicity:
\begin{compactdesc}
\item[\emph{evolution masking}:]
the monitor may not properly report the evolution of the system execution, e.g., mitigation of violation severity may not be detected;
\item[\emph{violation masking}:]
as a special case of \emph{evolution masking}, the first violation episode during the system execution ``masks'' the following ones.
\end{compactdesc}

The information not reported by \classicMon because of information masking, is very useful in several contexts. First of all, in some systems, the first violation of the specification does not mean that the system is not operating anymore, and one may want to continue monitoring and detect all the succeeding violations; this is the case, e.g., of the monitoring approach reported by Selyunin et al.~\cite{selyunin2017runtime} in which all the violations of the SENT protocol must be detected. Moreover, having a precise description of the system evolution is important for the usefulness of the monitoring; for example, the monitoring of the \speed in Fig.~\ref{fig:runningExampleIntro} could be used in a vehicle for checking the speed and notifying the driver whenever the speed is approaching the critical limit; if the monitor is not able to precisely capture the severity of violation, it cannot be used for this type of application.

Some works~\cite{selyunin2017runtime,cimatti2019assumption,onlineResetTCAD2022} try to mitigate the monotonicity issues, by ``resetting'' the monitor at specific points. A  recent approach has been proposed by Zhang et al.~\cite{onlineResetTCAD2022} (called \resetMon in the following) that is able to identify each ``violation episode'' (i.e., it solves the problem of \emph{violation masking}), but does not solve the \emph{evolution masking} problem. For the example in Fig.~\ref{fig:runningExampleIntro}, \resetMon is able to detect the two violation episodes in intervals $[20, 35]$ and $[40, 45]$, but it is not able to report that the speed decreases after $\timeSymbol = 10$ (in a non-violating situation), and that the severity of the violation is mitigated after $\timeSymbol = 30$.

\noindent{\bf Contribution.} In this paper, in order to provide more information about the evolution of the monitored system, we propose to monitor the \emph{causation} of  violation or satisfaction, instead of considering the robustness directly. To do this, we rely on the notion of \emph{epoch}~\cite{bartocci2018localizing}. At each instant, the \emph{violation (satisfaction) epoch} identifies the time instants at which the evaluation of the atomic propositions of the specification $\varphi$ causes the violation (satisfaction) of $\varphi$.

Based on the notion of epoch, we define a \emph{Boolean causation monitor} (called \boolCauseMon) that, at runtime, not only assesses the specification violation/satisfaction, but also tells whether each instant is relevant to it. Namely, \boolCauseMon marks each current instant \timeSymbol as
\begin{inparaenum}[(i)]
\item a \emph{violation causation instant}, if \timeSymbol is added to the violation epoch;
\item a \emph{satisfaction causation instant}, if \timeSymbol is added to the satisfaction epoch;
\item an \emph{irrelevant instant}, if \timeSymbol is not added to any epoch.
\end{inparaenum}
We show that \boolCauseMon is able to detect all the violation episodes (so solving the \emph{violation masking} issue), as violation causation instants can be followed by irrelevant instants. 
Moreover, we show that the information provided by the classic Boolean online monitor can be derived from that of the Boolean causation monitor \boolCauseMon.

However, \boolCauseMon just tells us whether the current instant is a (violation or satisfaction) causation instant or not, but does not report \emph{how far} the instant is from being a causation instant. To this aim, we introduce the notion of \emph{causation distance}, as a quantitative measure characterizing the spatial distance of the signal value at $b$ from turning $b$ into a causation instant. Then, we propose the \emph{quantitative causation monitor} (\quanCauseMon) that, at each instant, returns its causation distance. We show that using \quanCauseMon, besides solving the \emph{violation masking} problem, we also solve the \emph{evolution masking} problem. Moreover, we show that we can derive from \quanCauseMon both the  classic quantitative monitor \classicMon, and the Boolean causation monitor \boolCauseMon.

Experimental results show that the proposed monitors, not only provide more information, by they do it in an efficient way, not requiring a significant additional monitoring time w.r.t. the existing monitors.

\smallskip
\noindent{\bf Outline.}
\S{}\ref{sec:background} reports necessary background. We introduce \boolCauseMon in \S{}\ref{sec:booleanCausationMonitor}, and \quanCauseMon in \S{}\ref{sec:quantitativeCausationMonitor}. 
Experimental assessment of the two proposed monitors is reported in \S{}\ref{sec:experiment}. Finally, \S{}\ref{sec:relatedWork} discusses some related work, and \S{}\ref{sec:conclusions} concludes the paper.

\section{Preliminaries}\label{sec:background}
In this section, we review the fundamentals of \emph{signal temporal logic (STL)}
in \S{}\ref{sec:STLBasic}, and then introduce the existing classic online monitoring approach in \S{}\ref{sec:STLOnlineMonitor}.

\subsection{Signal Temporal Logic}\label{sec:STLBasic}
Let $T\in \Rpos$ be a positive real, and $\dimension\in\Npos$ be a positive integer. A \emph{$\dimension$-dimensional signal} is a function $\bw\colon [0,T]\to\R^\dimension$,
where $T$ is called the \emph{time horizon} of $\bw$. Given an arbitrary time instant $t\in[0, T]$, $\bw(t)$ is a $\dimension$-dimensional real vector; each dimension concerns a \emph{signal variable} that has a certain physical meaning, e.g., \speed, \rpm, \acceleration, etc. In this paper, we fix a set $\Var$ of variables and assume that a signal $\bw$ is \emph{spatially bounded}, i.e., for all $t\in[0, T]$, $\bw(t)\in\Omega$, where $\Omega$ is a $\dimension$-dimensional hyper-rectangle.

\emph{Signal temporal logic (STL)} is a widely-adopted specification language, used to describe the expected behavior of systems. In Def.~\ref{def:STLsyntax} and Def.~\ref{def:robSemantics}, we respectively review the syntax and the robust semantics of STL~\cite{maler2004monitoring,donze2010robust, fainekos2009robustness}.

\begin{mydefinition}[STL syntax]\label{def:STLsyntax}
In STL, the \emph{atomic propositions} $\alpha$ and the \emph{formulas} $\varphi$ are defined as follows:
\begin{align*}
 & \alpha 
\,::\equiv\,
 f(w_1, \dots, w_K) > 0 \qquad
& \varphi \,::\equiv\,
 \alpha \mid \bot
 \mid \neg \varphi 
 \mid \varphi \wedge \varphi 
 \mid \Box_{I}\varphi
 \mid \Diamond_{I}\varphi
 \mid \varphi \UntilOp{I} \varphi
\end{align*}
Here $f$ is a $K$-ary function $f:\R^K \to \R$, $w_1, \dots, w_K \in \Var$, and $I$ is a closed interval over $\Rnn$, i.e.,\ $I=[l,u]$, where $l,u \in \R$ and $l\le u$. In the case that $l = u$, we can use $l$ to stand for $I$. $\Box, \Diamond$ and $\mathcal{U}$ are temporal operators, which are known as \emph{always}, \emph{eventually} and \emph{until}, respectively. The always operator $\Box$ and eventually operator $\Diamond$ are two special cases of the until operator $\mathcal{U}$, where $\Diamond_{I}\varphi\equiv\top\UntilOp{I}\varphi$ and $\Box_{I}\varphi\equiv\lnot\Diamond_{I}\lnot\varphi$. Other common connectives such as $\lor, \rightarrow$ are introduced as syntactic sugar: $\varphi_1\lor\varphi_2\equiv \neg(\neg\varphi_1\land\neg\varphi_2)$, $\varphi_1\to\varphi_2 \equiv \neg\varphi_1 \lor \varphi_2$. 
\end{mydefinition}

\begin{mydefinition}[STL robust semantics]\label{def:robSemantics} 
Let $\bw$ be a signal, $\varphi$ be an STL formula and $\tau \in\Rpos$ be an instant. The \emph{robustness} $\Robust{\bw}{\varphi}{\tau} \in \R \cup \{\infty,-\infty\}$ of $\bw$ w.r.t. $\varphi$ at $\tau$ is defined by induction on the construction of formulas, as follows. 
\begin{align*}
&\Robust{\bw}{\alpha}{\tau} \;\Defeq\; f(\bw(\tau)) 
\qquad
\Robust{\bw}{\bot}{\tau} \;\Defeq\; -\infty
\qquad
\Robust{\bw}{\neg \varphi}{\tau} \;\Defeq\; -\Robust{\bw}{\varphi}{\tau}
\\
&\Robust{\bw}{\varphi_1\land \varphi_2}{\tau} \;\Defeq\; \min\left(\Robust{\bw}{\varphi_1}{\tau}, \Robust{\bw}{\varphi_2}{\tau}\right) 
\\
&\Robust{\bw}{\Box_{I}\varphi}{\tau} \;\Defeq\; \inf_{t\in \tau + I}{\Robust{\bw}{\varphi}{t}}
\qquad
\Robust{\bw}{\Diamond_{I}\varphi}{\tau} \;\Defeq\; \sup_{t\in \tau + I}{\Robust{\bw}{\varphi}{t}}
\\
&\Robust{\bw}{\varphi_1 \UntilOp{I} \varphi_2}{\tau} \;\Defeq\; \sup_{t \in \tau + I} \min\left(\Robust{\bw}{\varphi_2}{t}, \inf_{t' \in [\tau, t)} \Robust{\bw}{\varphi_1}{t'} \right)
\end{align*}
Here, $\tau + I$ denotes the interval $[l+\tau, u+\tau]$. 
\end{mydefinition}

The original STL semantics is Boolean, which represents whether a signal $\bw$ satisfies $\varphi$ at an instant $\tau$, i.e., whether $(\bw, \tau) \models \varphi$.
The robust semantics in Def.~\ref{def:robSemantics} is a quantitative extension that refines the original Boolean STL semantics, in the sense that, $\Robust{\bw}{\varphi}{\tau} > 0$ implies $(\bw, \tau)\models\varphi$, and $\Robust{\bw}{\varphi}{\tau} < 0$ implies $(\bw, \tau)\not\models\varphi$. More details can be found in~\cite[Prop. 16]{fainekos2009robustness}. 

\subsection{Classic Online Monitoring of STL}\label{sec:STLOnlineMonitor}

STL robust semantics in Def.~\ref{def:robSemantics} provides an offline monitoring approach for \emph{complete signals}. \emph{Online monitoring}, instead, targets a growing \emph{partial signal} at runtime.
Besides the verdicts $\top$ and $\bot$, an online monitor can also report the verdict \textunknown (denoted as $\unknown$), which represents a status when the satisfaction of the signal to $\varphi$ is not decided yet. In the following, we formally define partial signals and introduce online monitors for STL.

\medskip
Let $T$ be the time horizon of a signal $\bw$, and let $[a, b]\subseteq [0, T]$ be a sub-interval in the time domain $[0, T]$. A \emph{partial signal} $\bwp{a}{b}$ is a function which is only defined in the interval $[a, b]$; in the remaining domain $[0,T]\setminus[a, b]$, we denote that $\bwp{a}{b} = \epsilon$, where $\epsilon$ stands for a value that is not defined. 

Specifically, if $a = 0$ and $b\in(a, T]$, a partial signal $\bwp{a}{b}$ is called a \emph{prefix} (partial) signal; dually, if $b = T$ and $a\in[0, b)$, a partial signal $\bwp{a}{b}$ is called a \emph{suffix} (partial) signal.
Given a prefix signal \bwp{0}{b}, a \emph{completion} $\bwp{0}{b}\cdot\bwp{b}{T}$ of \bwp{0}{b} is defined as the concatenation of \bwp{0}{b} with a suffix signal \bwp{b}{T}.

\begin{mydefinition}[Classic Boolean STL online monitor]\label{def:boolClassicMonitor}
Let $\bwp{0}{b}$ be a prefix signal, and $\varphi$ be an STL formula. An online monitor $\CMon{\bwp{0}{b}}{\varphi}{\tau}$ returns a verdict in $\{\top, \bot, \unknown\}$ (namely, \true, \false, and \textunknown), as follows:
\begin{align*}
\CMon{\bwp{0}{b}}{\varphi}{\tau} \;\Defeq\;
\begin{cases}
\top & \text{if } \forall \bwp{b}{T}.\, \Robust{\bwp{0}{b}\cdot\bwp{b}{T}}{\varphi}{\tau} > 0 \\
\bot & \text{if } \forall \bwp{b}{T}.\, \Robust{\bwp{0}{b}\cdot\bwp{b}{T}}{\varphi}{\tau} < 0\\
\unknown & \text{otherwise}
\end{cases}
\end{align*}
\end{mydefinition} 
Namely, the verdicts of $\CMon{\bwp{0}{b}}{\varphi}{\tau}$ are interpreted as follows:
\begin{compactitem}
\item if any possible completion $\bwp{0}{b}\cdot\bwp{b}{T}$ of $\bwp{0}{b}$ satisfies $\varphi$, then \bwp{0}{b} satisfies $\varphi$;
\item if any possible completion $\bwp{0}{b}\cdot \bwp{b}{T}$ of $\bwp{0}{b}$ violates $\varphi$, then \bwp{0}{b} violates $\varphi$;
\item otherwise (i.e., there is a completion $\bwp{0}{b}\cdot\bwp{b}{T}$ that satisfies $\varphi$, and there is a completion $\bwp{0}{b}\cdot\bwp{b}{T}$ that violates $\varphi$), then $\CMon{\bwp{0}{b}}{\varphi}{\tau}$ reports \textunknown.
\end{compactitem}

\medskip
Note that, by Def.~\ref{def:boolClassicMonitor} only, we cannot synthesize a feasible online monitor, because the possible completions for $\bwp{0}{b}$ are infinitely many. A constructive online monitor is introduced in~\cite{deshmukh2017robust}, which implements the functionality of Def.~\ref{def:boolClassicMonitor} by computing the \emph{reachable} robustness of $\bwp{0}{b}$. We review this monitor in Def.~\ref{def:classicMonitor}.

\begin{mydefinition}[Classic Quantitative STL online monitor (\classicMon)]\label{def:classicMonitor}
    Let $\bwp{0}{b}$ be a prefix signal, and let $\varphi$ be an STL formula. We denote by $\Rmax$ and $\Rmin$ the possible \emph{maximum} and \emph{minimum bounds} of the robustness $\Robust{\bw}{\alpha}{\tau}$\footnote{$\Robust{\bw}{\alpha}{\tau}$ is bounded because $\bw$ is bounded by $\Omega$. In practice, if $\Omega$ is not know, we set $\Rmax$ and $\Rmin$ to, respectively, $\infty$ and $-\infty$.}. Then, an \emph{online monitor} 
    $\Mon{\bwp{0}{b}}{\varphi}{\tau}$, which returns a sub-interval  of $[\Rmin, \Rmax]$ at the instant $b$, is  defined as follows, by induction on the construction of formulas.
    \begin{align*}
        &\Mon{\bwp{0}{b}}{\alpha}{\tau} \;\Defeq\;  \begin{cases}
 \big[f\left(\bwp{0}{b}(\tau)\right), f\left(\bwp{0}{b}(\tau)\right)\big] & \text{if } \tau\in [0, b]\\
 \big[\Rmin, \Rmax\big]& \text{otherwise}
 \end{cases}\\
 &\Mon{\bwp{0}{b}}{\neg\varphi}{\tau} \;\Defeq\; - \Mon{\bwp{0}{b}}{\varphi}{\tau} \\
 &\Mon{\bwp{0}{b}}{\varphi_1\land\varphi_2}{\tau} \;\Defeq\; \min\Big(\Mon{\bwp{0}{b}}{\varphi_1}{\tau}, \Mon{\bwp{0}{b}}{\varphi_2}{\tau}\Big) \\
 &\Mon{\bwp{0}{b}}{\Box_I\varphi}{\tau} \;\Defeq\; \inf_{t\in \tau+I}\Big(\Mon{\bwp{0}{b}}{\varphi}{t}\Big) \\
 &\Mon{\bwp{0}{b}}{\varphi_1\UntilOp{I}\varphi_2}{\tau} \;\Defeq\; \sup_{t\in\tau +I}\min\Big(\Mon{\bwp{0}{b}}{\varphi_2}{t}, \inf_{t'\in[\tau, t)}\Mon{\bwp{0}{b}}{\varphi_1}{t'}\Big)
    \end{align*}
  Here, $f$ is defined as in Def.~\ref{def:STLsyntax}, and the arithmetic rules over intervals $I=[l, u]$ are defined as follows:
      $-I\Defeq[-u, -l] \text{ and } \min(I_1,I_2) \Defeq [\min(l_1, l_2), \min(u_1, u_2)]$.
\end{mydefinition}

We denote by $\MonU{\bwp{0}{b}}{\varphi}{\tau}$ and $\MonL{\bwp{0}{b}}{\varphi}{\tau}$ the upper bound and the lower bound of $\Mon{\bwp{0}{b}}{\varphi}{\tau}$ respectively. Intuitively, the two bounds together form the reachable robustness interval of the completion $\bwp{0}{b}\cdot\bwp{b}{T}$, under any possible suffix signal \bwp{b}{T}. For instance, in Fig.~\ref{fig:runningExample}, the upper bound $\MonUnoArgs$ at $b= 20$ is 0, which indicates that the robustness of the completion of the signal \speed, under any suffix, can never be larger than 0.

The quantitative online monitor \classicMon in Def.~\ref{def:classicMonitor} refines the Boolean one in Def.~\ref{def:boolClassicMonitor}, and the Boolean monitor can be derived from \classicMon as follows:
\begin{compactitem}
\item if $\MonL{\bwp{0}{b}}{\varphi}{\tau} > 0$, it implies that $\CMon{\bwp{0}{b}}{\varphi}{\tau} = \top$;
\item if $\MonU{\bwp{0}{b}}{\varphi}{\tau} < 0$, it implies that $\CMon{\bwp{0}{b}}{\varphi}{\tau} = \bot$;
\item otherwise, if $\MonL{\bwp{0}{b}}{\varphi}{\tau} < 0$ and $\MonU{\bwp{0}{b}}{\varphi}{\tau} > 0$, $\CMon{\bwp{0}{b}}{\varphi}{\tau} = \unknown$.
\end{compactitem}

\medskip
The classic online monitors are \emph{monotonic} by definition. In the Boolean monitor (Def.~\ref{def:boolClassicMonitor}), with the growth of $\bwp{0}{b}$, $\CMon{\bwp{0}{b}}{\varphi}{\tau}$ can only turn from \unknown to $\{\bot, \top\}$, but never the other way around. In the quantitative one (Def.~\ref{def:classicMonitor}), as shown in Lem.~\ref{lem:STLmono}, $\MonU{\bwp{0}{b}}{\varphi}{\tau}$ and $\MonL{\bwp{0}{b}}{\varphi}{\tau}$ are both monotonic, the former one decreasingly, the latter one increasingly. An example can be observed in Fig.~\ref{fig:runningExample}.

\begin{mylemma}[Monotonicity of STL online monitor]\label{lem:STLmono}
Let $\Mon{\bwp{0}{b}}{\varphi}{\tau}$ be the quantitative online monitor for a partial signal $\bwp{0}{b}$ and an STL formula $\varphi$. With the growth of the partial signal $\bwp{0}{b}$, the upper bound $\MonU{\bwp{0}{b}}{\varphi}{\tau}$ monotonically decreases, and the lower bound $\MonL{\bwp{0}{b}}{\varphi}{\tau}$ monotonically increases, i.e., for two time instants $b_1, b_2 \in [0, T]$, if $b_1 < b_2$, we have
\begin{inparaenum}[(i)]
    \item $\MonU{\bwp{0}{b_1}}{\varphi}{\tau} \geq \MonU{\bwp{0}{b_2}}{\varphi}{\tau}$, and
    \item $\MonL{\bwp{0}{b_1}}{\varphi}{\tau} \leq \MonL{\bwp{0}{b_2}}{\varphi}{\tau}$.
\end{inparaenum}
\end{mylemma}

\begin{proof}
This can be proved by induction on the structures of STL formulas. The detailed proof is given in Appendix~\ref{sec:monotonicityProof}. \qed
\end{proof}

\section{Boolean Causation Online Monitor}\label{sec:booleanCausationMonitor}

As explained in~\S{}\ref{sec:intro}, monotonicity of classic online monitors causes different types of \emph{information masking}, which prevents some information from being delivered.
In this section, we introduce a novel \emph{Boolean causation (online) monitor} \boolCauseMon, that solves the \emph{violation masking} issue (see \S{}\ref{sec:intro}). \boolCauseMon is defined based on \emph{online signal diagnostics}~\cite{onlineResetTCAD2022,bartocci2018localizing}, which reports the \emph{cause} of violation or satisfaction of the specification at the atomic proposition level. 

\begin{mydefinition}[Online signal diagnostics]\label{def:signalDiagnostics}
Let $\bwp{0}{b}$ be a partial signal and $\varphi$ be an STL specification. At an instant $b$, online signal diagnostics returns a \emph{violation epoch} $\DiagFSimp{\bwp{0}{b}}{\varphi}{\tau}$, under the condition $\MonU{\bwp{0}{b}}{\varphi}{\tau} < 0$, as follows:
    \begin{align*}
        &\DiagFSimp{\bwp{0}{b}}{\alpha}{\tau} \;\Defeq\; \begin{cases}
            \{\langle\alpha, \tau\rangle\} &\text{if } \MonU{\bwp{0}{b}}{\alpha}{\tau} < 0\\
            \emptyset & \text{otherwise}
        \end{cases} \\
         &\DiagFSimp{\bwp{0}{b}}{\neg\varphi}{\tau} \;\Defeq\; \DiagTSimp{\bwp{0}{b}}{\varphi}{\tau}\\ 
        &\DiagFSimp{\bwp{0}{b}}{\varphi_1\land\varphi_2}{\tau} \;\Defeq\; \bigcup_{\footnotesize\substack{i\in\{1, 2\} \text{ s.t.}\\\MonU{\bwp{0}{b}}{\varphi_i}{\tau}<0}}\hspace{-10pt}\DiagFSimp{\bwp{0}{b}}{\varphi_i}{\tau} \\
        &\DiagFSimp{\bwp{0}{b}}{\Box_I\varphi}{\tau} \;\Defeq\; \bigcup_{\footnotesize\substack{t\in \tau + I \text{ s.t.}\\ \MonU{\bwp{0}{b}}{\varphi}{t} < 0}}{\hspace{-10pt}\DiagFSimp{\bwp{0}{b}}{\varphi}{t}} \\
        &\DiagFSimp{\bwp{0}{b}}{\varphi_1\UntilOp{I}\varphi_2}{\tau} \;\Defeq\; \hspace{-12pt}\bigcup_{\footnotesize\substack{t\in \tau+I \text{ s.t.} \\ \MonU{\bwp{0}{b}}{\varphi_1\UntilOp{t}\varphi_2}{\tau}< 0}} \hspace{-8pt}\left(\DiagFSimp{\bwp{0}{b}}{\varphi_2}{t}\cup \bigcup_{t'\in [\tau, t)}\DiagFSimp{\bwp{0}{b}}{\varphi_1}{t'}\right)
    \end{align*}
and a \emph{satisfaction epoch} $\DiagTSimp{\bwp{0}{b}}{\varphi}{\tau}$, under the condition $\MonL{\bwp{0}{b}}{\varphi}{\tau}>0$, as follows:
\begin{align*}
    &\DiagTSimp{\bwp{0}{b}}{\alpha}{\tau} \;\Defeq\; \begin{cases}
            \{\langle\alpha, \tau\rangle\} &\text{if } \MonL{\bwp{0}{b}}{\alpha}{\tau} > 0\\
            \emptyset & \text{otherwise}
        \end{cases}\\
    &\DiagTSimp{\bwp{0}{b}}{\neg\varphi}{\tau} \;\Defeq\; \DiagFSimp{\bwp{0}{b}}{\varphi}{\tau}\\
    &\DiagTSimp{\bwp{0}{b}}{\varphi_1\land\varphi_2}{\tau} \;\Defeq\; \bigcup_{\footnotesize\substack{i\in\{1, 2\} \text{ s.t.}\\\MonL{\bwp{0}{b}}{\varphi_i}{\tau}>0}}\DiagTSimp{\bwp{0}{b}}{\varphi_i}{\tau}\\
    &\DiagTSimp{\bwp{0}{b}}{\Box_I\varphi}{\tau} \;\Defeq\; \bigcup_{\footnotesize\substack{t\in \tau + I \text{ s.t.}\\ \MonL{\bwp{0}{b}}{\varphi}{t} > 0}}{\DiagTSimp{\bwp{0}{b}}{\varphi}{t}}\\
    &\DiagTSimp{\bwp{0}{b}}{\varphi_1\UntilOp{I}\varphi_2}{\tau} \;\Defeq\; \hspace{-12pt}\bigcup_{\footnotesize\substack{t\in \tau+I \text{ s.t.} \\ \MonL{\bwp{0}{b}}{\varphi_1\UntilOp{t}\varphi_2}{\tau} > 0}} \hspace{-8pt}\left(\DiagTSimp{\bwp{0}{b}}{\varphi_2}{t}\cup \bigcup_{t'\in[\tau, t)}\DiagTSimp{\bwp{0}{b}}{\varphi_1}{t'}\right)
\end{align*}
If the conditions are not satisfied, $\DiagFSimp{\bwp{0}{b}}{\varphi}{\tau}$ and $\DiagTSimp{\bwp{0}{b}}{\varphi}{\tau}$ are both $\emptyset$. 
Note that the definition is recursive, thus the conditions should also be checked for computing the violation and satisfaction epochs of the sub-formulas of $\varphi$.

Computation for other operators can be inferred by the presented ones and the STL syntax (Def.~\ref{def:STLsyntax}).
\end{mydefinition}

Intuitively, when a partial signal $\bwp{0}{b}$ violates a specification $\varphi$, a violation epoch starts collecting the evaluations (identified by pairs of atomic propositions and instants) of the signal at the atomic proposition level, that cause the violation of the whole formula $\varphi$ (which also applies to the satisfaction cases in a dual manner). This is done inductively, based on the semantics of different operators:
\begin{compactitem}
    \item in the case of an atomic proposition $\alpha$, if $\alpha$ is violated at $\tau$, it collects $\langle\alpha, \tau\rangle$;
    \item in the case of a negation $\neg\varphi$, it collects the satisfaction epoch of $\varphi$;
    \item in the case of a conjunction $\varphi_1\land\varphi_2$, it collects the union of the violation epochs of the sub-formulas violated by the partial signal;
    \item in the case of an \emph{always} operator $\Box_I\varphi$, it collects the epochs of the sub-formula $\varphi$ at all the instants $t$ where $\varphi$ is evaluated as being violated.
    \item in the case of an \emph{until} operator $\varphi_1 \UntilOp{I} \varphi_2$, it collects the epochs of the sub-formula $\varphi_2$ at all the instants $t$ and the epochs of $\varphi_1$ at the instants $t'\in [\tau,t)$, in the case where the clause ``$\varphi_1$ until $\varphi_2$'' is violated at $t$. 
\end{compactitem}

\begin{figure}[!tb]
    \begin{tabular}{ccc}
    \multirow[t]{2}{*}{\begin{minipage}{0.39\textwidth}
        \vspace{+28pt}\centering\includegraphics[width=0.7\linewidth]{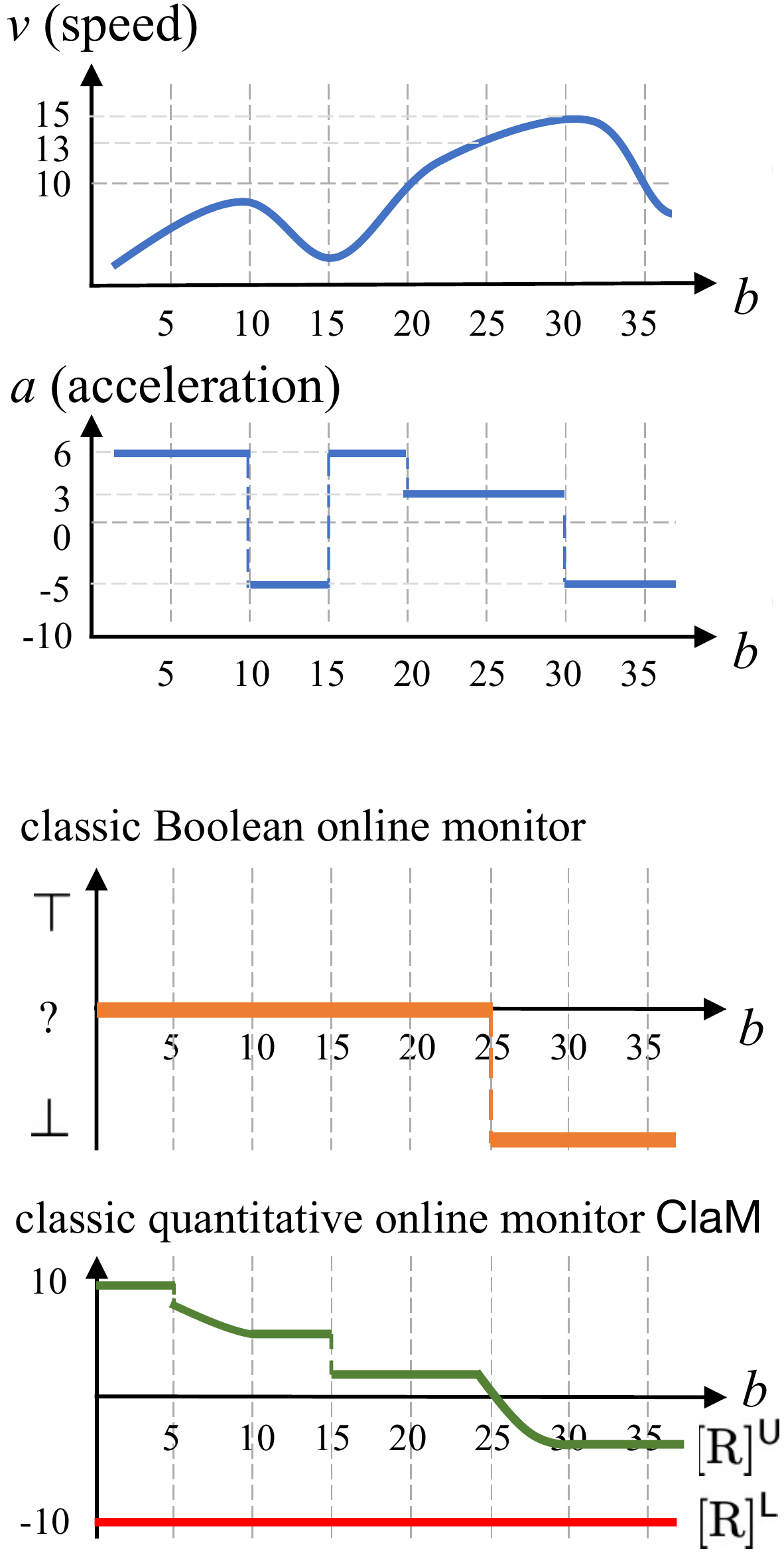}
        \caption{Classic monitor (\classicMon) result for the STL specification: \footnotesize{$\Box_{[0,100]}(v>10 \to \Diamond_{[0,5]}(a<0))$}}
        \label{fig:runningExample}
      \end{minipage}}  & $\;$ &
      \begin{minipage}[t]{0.58\textwidth}
        \centering
        \includegraphics[width=\linewidth]{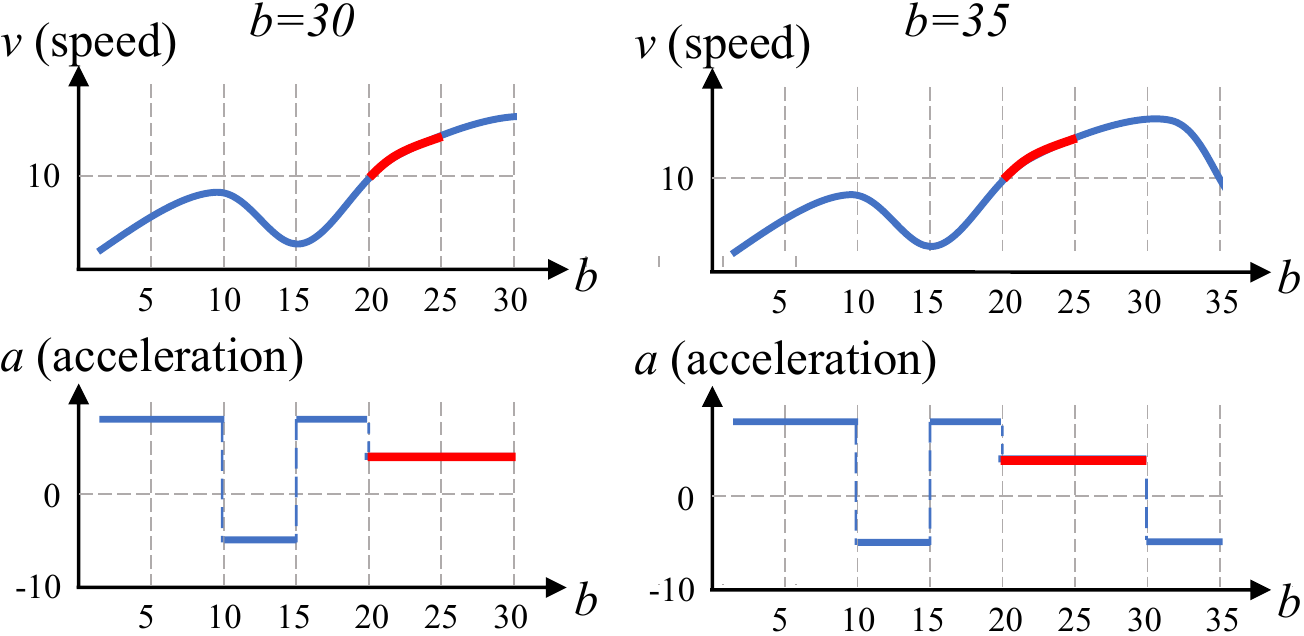}
        \caption{The violation epochs (the red parts) respectively when $b= 30$ and $b=35$}
        \label{fig:violationEpochs}
      \end{minipage} \vspace{15pt} \\
      &&\begin{minipage}[t]{0.6\textwidth}
        \centering
        \includegraphics[width=0.7\linewidth]{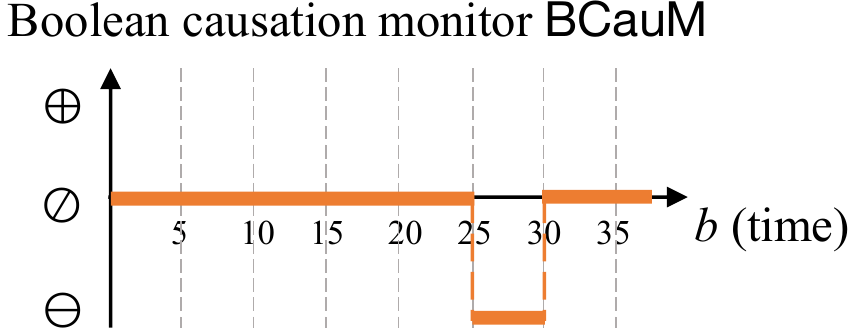}
        \caption{Boolean causation monitor (\boolCauseMon) result}
        \label{fig:boolFineMonitor}
      \end{minipage} 
    \end{tabular}
  \end{figure}
  
\begin{myexample}\label{eg:epochs}
The example in Fig.~\ref{fig:runningExample} illustrates how an epoch is collected. The specification requires that whenever the \speed is higher than 10, the car should decelerate within 5 time units. As shown by the classic monitor, the specification is violated at $b = 25$, since $v$ becomes higher than 10 at $20$ but $a$ remains positive during $[20, 25]$.
Note that the specification can be rewritten as $\varphi \equiv \Box_{[0,100]}(\neg(v>10)\lor \Diamond_{[0,5]}(a<0))$. For convenience, we name the sub-formulas of $\varphi$ as follows: 
    \begin{gather*}
        \varphi' \equiv \neg(v>10)\lor \Diamond_{[0,5]}(a<0) \qquad \varphi_1\equiv \neg(v>10) \qquad \varphi_2\equiv \Diamond_{[0,5]}(a<0) \\ \alpha_1 \equiv v >10 \qquad \alpha_2 \equiv a< 0
    \end{gather*}
    Fig.~\ref{fig:violationEpochs} shows the violation epochs at two instants 30 and 35. First, at $b = 30$, 
    \begin{align*}
        \DiagFSimp{\bwp{0}{30}}{\varphi}{0} &= \big(\textstyle\bigcup_{t\in[20, 25]}\DiagTSimp{\bwp{0}{30}}{\alpha_1}{t}\big) \cup \big(\textstyle\bigcup_{t\in[20,30]}\DiagFSimp{\bwp{0}{30}}{\alpha_2}{t}\big) \\
        &= \langle\alpha_1, [20, 25]\rangle \cup \langle\alpha_2, [20, 30]\rangle
    \end{align*}
Similarly, the violation epoch $\DiagFSimp{\bwp{0}{35}}{\varphi}{0}$ at $b=35$ is the same as that at $b = 30$. Intuitively, the epoch at $b = 30$ shows the cause of the violation of $\bwp{0}{30}$; then since signal $a<0$ in $[30,35]$, this segment is not considered as the cause of the violation, so the epoch remains the same at $b = 35$.
\hfill $\lhd$
\end{myexample}

\begin{mydefinition}[Boolean causation monitor (\boolCauseMon)] \label{def:boolFineMonitor}
Let $\bwp{0}{b}$ be a partial signal and $\varphi$ be an STL specification. We denote by $\alphaset$ the set of atomic propositions of $\varphi$.  At each instant $b$, a \emph{Boolean causation (online) monitor} \boolCauseMon returns a verdict in $\{\negCause, \posCause, \nothing\}$ (called \emph{violation causation}, \emph{satisfaction causation} and \emph{irrelevant}), which is defined as follows, 
    \begin{align*}
        \CDiag{\bwp{0}{b}}{\varphi}{\tau} \;\Defeq\;
        \begin{cases}
            \negCause & \text{if }  \exists\alpha\in\alphaset.\;\langle\alpha, b\rangle\in\DiagFSimp{\bwp{0}{b}}{\varphi}{\tau} \\
            \posCause & \text{if }\exists\alpha\in\alphaset.\;\langle\alpha, b\rangle\in \DiagTSimp{\bwp{0}{b}}{\varphi}{\tau} \\
            \nothing & \text{otherwise} 
        \end{cases} 
    \end{align*}
    An instant $b$ is called a \emph{violation/satisfaction causation instant} if $\CDiag{\bwp{0}{b}}{\varphi}{\tau}$ returns $\negCause$/$\posCause$, or an \emph{irrelevant instant} if $\CDiag{\bwp{0}{b}}{\varphi}{\tau}$ returns $\nothing$.
\end{mydefinition}

Intuitively, if the current instant $b$ (with the related $\alpha$) is included in the epoch (thus the signal value at $b$ is relevant to the violation/satisfaction of $\varphi$), \boolCauseMon will report a \emph{violation/satisfaction causation} (\negCause/\posCause); otherwise, it will report \emph{irrelevant} (\nothing). Notably \boolCauseMon is non-monotonic, in that even if it reports $\negCause$ or $\posCause$ at some instant $b$, it may still report $\nothing$  after $b$. This feature allows \boolCauseMon to bring more information, e.g., it can detect the end of a violation episode and the start of a new one (i.e., it solves the {\it violation masking} issue in \S{}\ref{sec:intro}); see Ex.~\ref{eg:boolCauseMonitor}.

\begin{myexample}\label{eg:boolCauseMonitor}
Based on the signal diagnostics  in Fig.~\ref{fig:violationEpochs}, the Boolean causation monitor \boolCauseMon reports the result shown as in Fig.~\ref{fig:boolFineMonitor}.  

Compared to the classic Boolean monitor in Fig.~\ref{fig:runningExample}, \boolCauseMon brings more information, in the sense that it detects the end of the violation episode at $\timeSymbol = 30$, by going from $\negCause$ to $\nothing$, when the signal $a$ becomes negative. \hfill $\lhd$
\end{myexample}

Thm.~\ref{lem:boolFineMonitor} states the relation of \boolCauseMon with the classic Boolean online monitor.

\begin{mytheorem}\label{lem:boolFineMonitor}
The Boolean causation monitor \boolCauseMon in Def.~\ref{def:boolFineMonitor} refines the classic Boolean online monitor in Def.~\ref{def:boolClassicMonitor}, in the following sense:
    \begin{compactitem}
        \item $\CMon{\bwp{0}{b}}{\varphi}{\tau} = \bot  \quad\textit{iff.}\quad \bigvee_{t\in[0,b]}\left(\CDiag{\bwp{0}{t}}{\varphi}{\tau} = \negCause\right)$
        \item $\CMon{\bwp{0}{b}}{\varphi}{\tau} = \top  \quad\textit{iff.}\quad \bigvee_{t\in[0,b]}\left(\CDiag{\bwp{0}{t}}{\varphi}{\tau} = \posCause\right)$
        \item $\CMon{\bwp{0}{b}}{\varphi}{\tau} = \,\unknown\,  \quad\textit{iff.}\quad \bigwedge_{t\in[0,b]}\left(\CDiag{\bwp{0}{t}}{\varphi}{\tau} = \nothing\right)$
    \end{compactitem}
\end{mytheorem}

\begin{proof}
The proof is based on Defs.~\ref{def:signalDiagnostics} and \ref{def:boolFineMonitor}, Lem.~\ref{lem:STLmono} about the monotonicity of classic STL online monitors, and Lems.~\ref{lem:notEmpty} and \ref{lem:firstVio} in Appendix~\ref{sec:boolCauseMonProof} about the properties of epochs. See the complete proof in Appendix~\ref{sec:boolCauseMonProof}. \qed
\end{proof}

\section{Quantitative Causation Online Monitor}\label{sec:quantitativeCausationMonitor}

Although \boolCauseMon in~\S\ref{sec:booleanCausationMonitor} is able to solve the {\it violation masking} issue, it still does not provide enough information about the evolution of the system signals, i.e., it does not solve the {\it evolution masking} issue introduced in \S{}\ref{sec:intro}.
To tackle this issue, we propose a \emph{quantitative (online) causation monitor} \quanCauseMon in Def.~\ref{def:quanFineMonitor}, which is a quantitative extension of \boolCauseMon. Given a partial signal $\bwp{0}{b}$, \quanCauseMon reports a \emph{violation causation distance} $\InsRobVio{\bwp{0}{b}}{\varphi}{\tau}$ and a \emph{satisfaction causation distance} $\InsRobSat{\bwp{0}{b}}{\varphi}{\tau}$, which, respectively, indicate \emph{how far} the signal value at the current instant $\timeSymbol$ is from turning $\timeSymbol$ into a violation causation instant and from turning $\timeSymbol$ into a satisfaction causation instant.

\begin{mydefinition}[Quantitative causation monitor (\quanCauseMon)]
\label{def:quanFineMonitor}
Let $\bwp{0}{b}$ be a partial signal, and $\varphi$ be an STL specification. At instant $b$, the quantitative causation monitor \quanCauseMon returns a \emph{violation causation distance} $\InsRobVio{\bwp{0}{b}}{\varphi}{\tau}$, as follows:
\begin{align*}
    &\InsRobVio{\bwp{0}{b}}{\alpha}{\tau} \;\Defeq\; \begin{cases}
        f(\bwp{0}{b}(\tau)) & \text{if } b = \tau\\
        \Rmax & \text{otherwise}
    \end{cases}\\
    &\InsRobVio{\bwp{0}{b}}{\neg\varphi}{\tau}\;\Defeq\; -\InsRobSat{\bwp{0}{b}}{\varphi}{\tau} \\
    &\InsRobVio{\bwp{0}{b}}{\varphi_1\land \varphi_2}{\tau} \;\Defeq\; \min\left(\InsRobVio{\bwp{0}{b}}{\varphi_1}{\tau}, \InsRobVio{\bwp{0}{b}}{\varphi_2}{\tau}\right) \\
    &\InsRobVio{\bwp{0}{b}}{\varphi_1\lor\varphi_2}{\tau} \;\Defeq\; \min\left(
    \begin{array}{l}
         \max\left(
         \InsRobVio{\bwp{0}{b}}{\varphi_1}{\tau}, \MonU{\bwp{0}{b}}{\varphi_2}{\tau}
         \right),  \\
        \max\left(
        \MonU{\bwp{0}{b}}{\varphi_1}{\tau}, \InsRobVio{\bwp{0}{b}}{\varphi_2}{\tau}
        \right)
    \end{array}
    \right)\\
    &\InsRobVio{\bwp{0}{b}}{\Box_I\varphi}{\tau}\;\Defeq\; \inf_{t\in\tau+I}\left(\InsRobVio{\bwp{0}{b}}{\varphi}{t}\right)\\
    &\InsRobVio{\bwp{0}{b}}{\Diamond_I\varphi}{\tau}\;\Defeq\; 
\inf_{t\in\tau+I}\left(
\max\left(
\InsRobVio{\bwp{0}{b}}{\varphi}{t}, \MonU{\bwp{0}{b}}{\Diamond_I\varphi}{\tau}
\right)
\right)
\\
    &\InsRobVio{\bwp{0}{b}}{\varphi_1\UntilOp{I}\varphi_2}{\tau}\;\Defeq\; 
    \inf_{t\in\tau+I}\left(
    \max\left(
    \begin{array}{l}
         \min\left(
         \begin{array}{l}
              \displaystyle\inf_{t'\in[\tau,t)}\InsRobVio{\bwp{0}{b}}{\varphi_1}{t'}  \\
               \InsRobVio{\bwp{0}{b}}{\varphi_2}{t}
         \end{array}
         \right)  \\
          \MonU{\bwp{0}{b}}{\varphi_1\UntilOp{I}\varphi_2}{\tau}
    \end{array}
    \right)
    \right)
\end{align*}
and a \emph{satisfaction causation distance} $\InsRobSat{\bwp{0}{b}}{\varphi}{\tau}$, as follows:
\begin{align*}
    &\InsRobSat{\bwp{0}{b}}{\alpha}{\tau} \;\Defeq\; \begin{cases}
        f(\bwp{0}{b}(\tau)) & \text{if } b = \tau\\
        \Rmin & \text{otherwise} 
    \end{cases}\\
    &\InsRobSat{\bwp{0}{b}}{\neg\varphi}{\tau}\;\Defeq\; - \InsRobVio{\bwp{0}{b}}{\varphi}{\tau} \\
    &\InsRobSat{\bwp{0}{b}}{\varphi_1\land\varphi_2}{\tau} \;\Defeq\; 
    \max\left(
    \begin{array}{l}
         \min\left(
         \InsRobSat{\bwp{0}{b}}{\varphi_1}{\tau}, \MonL{\bwp{0}{b}}{\varphi_2}{\tau}
         \right),  \\
        \min\left(
        \MonL{\bwp{0}{b}}{\varphi_1}{\tau}, \InsRobSat{\bwp{0}{b}}{\varphi_2}{\tau}
        \right)
    \end{array}
    \right)\\
    &\InsRobSat{\bwp{0}{b}}{\varphi_1\lor\varphi_2}{\tau}\;\Defeq\; \max\left(\InsRobSat{\bwp{0}{b}}{\varphi_1}{\tau}, \InsRobSat{\bwp{0}{b}}{\varphi_2}{\tau}\right)\\
    &\InsRobSat{\bwp{0}{b}}{\Box_I\varphi}{\tau} \;\Defeq\; 
\sup_{t\in\tau+I}\left(
\min\left(
\InsRobSat{\bwp{0}{b}}{\varphi}{t}, \MonL{\bwp{0}{b}}{\Box_I\varphi}{\tau}
\right)
\right)
    \\
    &\InsRobSat{\bwp{0}{b}}{\Diamond_I\varphi}{\tau} \;\Defeq\; \sup_{t\in\tau+I}\left(\InsRobSat{\bwp{0}{b}}{\varphi}{t}\right)
    \\
     &\InsRobSat{\bwp{0}{b}}{\varphi_1\UntilOp{I}\varphi_2}{\tau} \Defeq
    \sup_{t\in\tau+I}\left(
    \max\left(
    \begin{array}{l}
         \min\left(
         \begin{array}{l}
           \displaystyle\sup_{t'\in[\tau, t)}\InsRobSat{\bwp{0}{b}}{\varphi_1}{t'}     \\
           \displaystyle\inf_{t'\in[\tau, t)}\MonL{\bwp{0}{b}}{\varphi_1}{t'}\\
           \MonL{\bwp{0}{b}}{\varphi_2}{t}
         \end{array}
         \right)  \\
         \min\left(
         \begin{array}{l}
              \displaystyle\inf_{t'\in[\tau, t)}\MonL{\bwp{0}{b}}{\varphi_1}{t'}  \\
            \InsRobSat{\bwp{0}{b}}{\varphi_2}{t}
         \end{array}
         \right) 
    \end{array}
    \right)
    \right)
\end{align*}
\end{mydefinition}
Intuitively, a violation causation distance $\InsRobVio{\bwp{0}{b}}{\varphi}{\tau}$ is the spatial distance of the  signal value $\bwp{0}{b}(b)$, at the current instant $b$, from turning $b$ into a violation causation instant such that $b$ is relevant to the violation of $\varphi$ (also applied to the satisfaction case dually). It is computed inductively on the structure of $\varphi$:
\begin{compactitem}
    \item Case atomic propositions $\alpha$: if $b = \tau$ (i.e., at which instant $\alpha$ should be evaluated), then the distance of $b$ from being a violation causation instant is $f(\bwp{0}{b}(b))$; otherwise, if $b\neq\tau$, despite the value of $f(\bwp{0}{b}(b))$, $b$ can never be a violation causation instant, according to Def.~\ref{def:signalDiagnostics}, because only $f(\bwp{0}{b}(\tau))$ is relevant to the violation of $\alpha$. Hence, the distance will be $\Rmax$;
    \item Case $\neg\varphi$: $b$ is a violation causation instant for $\neg\varphi$ if $b$ is a satisfaction causation instant for $\varphi$, so $\InsRobVio{\bwp{0}{b}}{\neg\varphi}{\tau}$ depends on $\InsRobSat{\bwp{0}{b}}{\varphi}{\tau}$;
    \item Case $\varphi_1\land\varphi_2$: $b$ is a violation causation instant for $\varphi_1\land\varphi_2$ if $b$ is a violation causation instant for either $\varphi_1$ or $\varphi_2$, so  $\InsRobVio{\bwp{0}{b}}{\varphi_1\land\varphi_2}{\tau}$ depends on the minimum between $\InsRobVio{\bwp{0}{b}}{\varphi_1}{\tau}$ and $\InsRobVio{\bwp{0}{b}}{\varphi_2}{\tau}$;
    \item Case $\varphi_1\lor\varphi_2$: $b$ is a violation causation instant for $\varphi_1\lor\varphi_2$ if, first, $\varphi_1\lor\varphi_2$ has been violated at $b$, and second, $b$ is the violation causation instant for either $\varphi_1$ or $\varphi_2$. Hence, $\InsRobVio{\bwp{0}{b}}{\varphi_1\lor\varphi_2}{\tau}$ depend on both the violation status (measured by $\MonU{\bwp{0}{b}}{\varphi_i}{\tau}$) of one sub-formula and the violation causation distance of the other sub-formula;
    \item Case $\Box_I\varphi$: $b$ is a violation causation instant for $\Box_I\varphi$ if $b$ is the violation causation instant for the sub-formula $\varphi$ evaluated at any instant in $\tau+I$. So, $\InsRobVio{\bwp{0}{b}}{\Box_I\varphi}{\tau}$ depends on the infimum of the violation causation distances regarding $\varphi$ evaluated at the instants in $\tau +I$;
    \item Case $\Diamond_I\varphi$: $b$ is a violation causation instant for $\Diamond_I\varphi$ if, first, $\Diamond_I\varphi$ has been violated at $b$, and second, $b$ is a violation causation instant for the sub-formula $\varphi$ evaluated at any instant in $\tau+I$. So, $\InsRobVio{\bwp{0}{b}}{\Diamond_I\varphi}{\tau}$ depends on both the violation status of $\Diamond_I\varphi$ (measured by $\MonU{\bwp{0}{b}}{\Diamond_I\varphi}{\tau}$) and the infimum of the violation causation distances of $\varphi$ evaluated in $\tau+I$.
    \item Case $\varphi_1\UntilOp{I}\varphi_2$: $\InsRobVio{\bwp{0}{b}}{\varphi_1\UntilOp{I}\varphi_2}{\tau}$ depends on, first, the violation status of the whole formula (measured by $\MonU{\bwp{0}{b}}{\varphi_1\UntilOp{I}\varphi_2}{\tau}$), and also, the infimum of the violation causation distances regarding the evaluation of ``$\varphi_1$ holds until $\varphi_2$'' at each instant in $\tau+I$. 
\end{compactitem}

\begin{figure}[!t]
\centering
\includegraphics[scale=0.8]{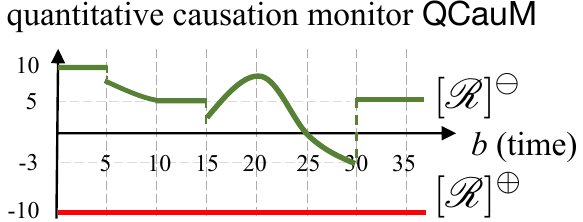}
\caption{Quantitative causation monitor (\quanCauseMon) result for Ex.~\ref{eg:epochs}}
\label{fig:quanMonitorEg}
\end{figure}

\begin{myexample}\label{eg:quanCauseMonitor}
Consider the quantitative causation monitor for the signals in Ex.~\ref{eg:epochs}. At $b = 30$, the violation causation distance is computed as:
{\mathcompact
\footnotesize
\begin{align*}
    &\InsRobVio{\bwp{0}{30}}{\varphi}{0} = \inf_{t\in[0,100]}\InsRobVio{\bwp{0}{30}}{\varphi'}{t} \\
    =&\inf_{t\in[0,100]}\left(\min\left(
    \begin{array}{l}
         \max\left(
            \InsRobVio{\bwp{0}{30}}{\varphi_1}{t},    
            \MonU{\bwp{0}{30}}{\varphi_2}{t}
         \right), \\
          \max\left(
               \MonU{\bwp{0}{30}}{\varphi_1}{t},  
                \InsRobVio{\bwp{0}{30}}{\varphi_2}{t}
          \right)    
    \end{array}
    \right)\right)
    \\
    =& \inf_{t\in[0,100]}\left(\min\left(
    \begin{array}{l}
         \max\left(-\InsRobSat{\bwp{0}{30}}{\alpha_1}{t}, \displaystyle\sup_{t'\in t+[0,5]}\MonU{\bwp{0}{30}}{\alpha_2}{t'}\right)  \\
         \max\left(-\MonL{\bwp{0}{30}}{\alpha_1}{t}, \max\left(
         \begin{array}{l}
         \MonU{\bwp{0}{30}}{\varphi_2}{t},
               \\
        \displaystyle\inf_{t'\in t+[0,5]}\InsRobVio{\bwp{0}{30}}{\alpha_2}{t'}  
         \end{array}
         \right) \right)
    \end{array}
    \right)\right)\\
    =& \max\left(-\MonL{\bwp{0}{30}}{\alpha_1}{25}, \MonU{\bwp{0}{30}}{\varphi_2}{25}, \inf_{t'\in[25, 30]}\InsRobVio{\bwp{0}{30}}{\alpha_2}{t'}\right)\\
    =&\max(-3, -3, -5) = -3
\end{align*}}

Similarly, at $b= 35$, the violation causation distance $\InsRobVio{\bwp{0}{35}}{\varphi}{0} = 5$. See the  result of \quanCauseMon shown in Fig.~\ref{fig:quanMonitorEg}. 
Compared to \classicMon in Fig.~\ref{fig:runningExample}, it is evident that \quanCauseMon provides much more information about the system evolution, e.g., it can report that, in the interval $[15, 20]$, the system satisfies the specification ``more'', as the \speed decreases. \hfill $\lhd$
\end{myexample}

By using the violation and satisfaction causation distances reported by \quanCauseMon jointly, we can infer the verdict of \boolCauseMon, as indicated by Thm.~\ref{theo:extension}.

\begin{mytheorem}\label{theo:extension}
The quantitative causation monitor \quanCauseMon
in Def.~\ref{def:quanFineMonitor} refines the Boolean causation monitor \boolCauseMon in Def.~\ref{def:boolFineMonitor}, in the sense that: 
\begin{compactitem}
    \item if $\InsRobVio{\bwp{0}{b}}{\varphi}{\tau} < 0$, it implies  $\CDiag{\bwp{0}{b}}{\varphi}{\tau} = \negCause$;
    \item if $\InsRobSat{\bwp{0}{b}}{\varphi}{\tau} > 0$, it implies  $\CDiag{\bwp{0}{b}}{\varphi}{\tau} = \posCause$;
    \item if $\InsRobVio{\bwp{0}{b}}{\varphi}{\tau} > 0$ and $\InsRobSat{\bwp{0}{b}}{\varphi}{\tau} < 0$, it implies $\CDiag{\bwp{0}{b}}{\varphi}{\tau} = \nothing$.
\end{compactitem}
\end{mytheorem}

\begin{proof}
The proof is generally based on mathematical induction. First, by Def.~\ref{def:quanFineMonitor} and Def.~\ref{def:signalDiagnostics}, it is straightforward that Thm.~\ref{theo:extension} holds for the atomic propositions.

Then, assuming that Thm.~\ref{theo:extension} holds for an arbitrary formula $\varphi$, we prove that Thm.~\ref{theo:extension} also holds for the composite formula $\varphi'$ constructed by applying STL operators to $\varphi$. The complete proof for all three cases is shown in Appendix~\ref{sec:theo1proof}.

As an instance, we show the proof for the first case with $\varphi'=\varphi_1 \vee \varphi_2$, i.e., we prove that $\InsRobVio{\bwp{0}{b}}{\varphi_1\lor\varphi_2}{\tau} < 0$ implies $\CDiag{\bwp{0}{b}}{\varphi_1\lor\varphi_2}{\tau} = \negCause$.
\begin{align*}
    &\InsRobVio{\bwp{0}{b}}{\varphi_1\lor\varphi_2}{\tau} < 0 \\
    \Rightarrow & \max\left(\InsRobVio{\bwp{0}{b}}{\varphi_1}{\tau}, \MonU{\bwp{0}{b}}{\varphi_2}{\tau}\right)< 0  &&\mycomment{by Def.~\ref{def:quanFineMonitor} and w.l.o.g.}\\
    \Rightarrow & \InsRobVio{\bwp{0}{b}}{\varphi_1}{\tau} < 0  && \mycomment{by def. of $\max$} \\
    \Rightarrow & \CDiag{\bwp{0}{b}}{\varphi_1}{\tau} = \negCause && \mycomment{by assumption}\\
    \Rightarrow & \DiagFSimp{\bwp{0}{b}}{\varphi_1\lor\varphi_2}{\tau} \supseteq \DiagFSimp{\bwp{0}{b}}{\varphi_1}{\tau} && \mycomment{by Def.~\ref{def:signalDiagnostics} and Thm.~\ref{lem:boolFineMonitor}}\\
    \Rightarrow & \exists \alpha.\; \langle \alpha, b\rangle\in \DiagFSimp{\bwp{0}{b}}{\varphi_1\lor\varphi_2}{\tau} &&\mycomment{by def. of $\supseteq$} \\
    \Rightarrow & \CDiag{\bwp{0}{b}}{\varphi_1\lor\varphi_2}{\tau} = \negCause &&\mycomment{by Def.~\ref{def:boolFineMonitor}} \tag*{\qed}
\end{align*}
\end{proof}

The relation between the quantitative causation monitor \quanCauseMon and the Boolean causation monitor \boolCauseMon, disclosed by Thm.~\ref{theo:extension}, can be visualized by the comparison between Fig.~\ref{fig:quanMonitorEg} and Fig.~\ref{fig:boolFineMonitor}. Indeed, when the violation causation distance reported by \quanCauseMon is negative in Fig.~\ref{fig:quanMonitorEg}, \boolCauseMon reports $\negCause$ in Fig.~\ref{fig:boolFineMonitor}. 

\medskip
Next, we present Thm.~\ref{theo:reconstruct}, which states the relation between the quantitative causation monitor \quanCauseMon and the classic quantitative monitor \classicMon. 

\begin{mytheorem}\label{theo:reconstruct}
The quantitative causation monitor \quanCauseMon in Def.~\ref{def:quanFineMonitor} refines the classic quantitative online monitor \classicMon in Def.~\ref{def:classicMonitor}, in the sense that, the monitoring results of \classicMon can be reconstructed from the results of \quanCauseMon, as follows:
\begin{align}
    &\MonU{\bwp{0}{b}}{\varphi}{\tau} = \inf_{t\in[0, b]}\InsRobVio{\bwp{0}{t}}{\varphi}{\tau} \label{eq:theo1case1}\\
    &\MonL{\bwp{0}{b}}{\varphi}{\tau} = \sup_{t\in[0, b]}\InsRobSat{\bwp{0}{t}}{\varphi}{\tau} \label{eq:theo1case2}
\end{align}
 
\end{mytheorem}

\begin{proof}
The proof is generally based on mathematical induction.
First, by Def.~\ref{def:quanFineMonitor} and Def.~\ref{def:classicMonitor}, it is straightforward that Thm.~\ref{theo:reconstruct} holds for the atomic propositions.

Then, we make the global assumption that Thm.~\ref{theo:reconstruct} holds for an arbitrary formula $\varphi$, i.e., both the two cases $\inf_{t\in[0,b]}\InsRobVio{\bwp{0}{t}}{\varphi}{\tau} = \MonU{\bwp{0}{b}}{\varphi}{\tau}$ and $\sup_{t\in[0,b]}\InsRobSat{\bwp{0}{t}}{\varphi}{\tau} = \MonL{\bwp{0}{b}}{\varphi}{\tau}$ hold. Based on this assumption, we prove that Thm.~\ref{theo:reconstruct} also holds for the composite formula $\varphi'$ constructed by applying STL operators to $\varphi$. 

As an instance, we prove $\inf_{t\in[0,b]}\InsRobVio{\bwp{0}{t}}{\varphi'}{\tau} = \MonU{\bwp{0}{b}}{\varphi'}{\tau}$ with $\varphi'=\varphi_1\lor\varphi_2$ as follows. The complete proof is presented in Appendix~\ref{sec:theo2proof}. 
        \begin{compactitem}
            \item First, if $b = \tau$, it holds that: 
            {\mathcompact
            \begin{align*}
                &\inf_{t\in[0,b]}\InsRobVio{\bwp{0}{t}}{\varphi_1\lor\varphi_2}{\tau}
                = \InsRobVio{\bwp{0}{\tau}}{\varphi_1\lor\varphi_2}{\tau} &&
                \\
                 =& \max\left(\MonU{\bwp{0}{\tau}}{\varphi_1}{\tau}, \MonU{\bwp{0}{\tau}}{\varphi_2}{\tau}\right) && \text{\footnotesize(by Def.~\ref{def:quanFineMonitor} and global assump.)}
                \\
                =& \MonU{\bwp{0}{b}}{\varphi_1\lor\varphi_2}{\tau} &&\text{\footnotesize(by Def.~\ref{def:classicMonitor})}
            \end{align*}
            }
            \item Then, we make a local assumption that, given an arbitrary $b$, it holds that $\inf_{t\in[0,b]}\InsRobVio{\bwp{0}{t}}{\varphi_1\lor\varphi_2}{\tau} = \MonU{\bwp{0}{b}}{\varphi_1\lor\varphi_2}{\tau}$. We prove that, for $b'$ which is the next sampling point to $b$, it holds that,
            {\mathcompact
           \begin{align*}
& \inf_{t\in[0,b']}\InsRobVio{\bwp{0}{t}}{\varphi_1\lor\varphi_2}{\tau} \\  = &\min\left(\MonU{\bwp{0}{b}}{\varphi_1\lor\varphi_2}{\tau}, \InsRobVio{\bwp{0}{b'}}{\varphi_1\lor\varphi_2}{\tau}\right)  & &\text{\footnotesize (by local assump.)} \\
                = &\min\left(
                \begin{array}{l}
                     \max\left(\MonU{\bwp{0}{b}}{\varphi_1}{\tau}, \MonU{\bwp{0}{b}}{\varphi_2}{\tau}\right),  \\
                     \max\left(\InsRobVio{\bwp{0}{b'}}{\varphi_1}{\tau}, \MonU{\bwp{0}{b'}}{\varphi_2}{\tau}\right), \\
                     \max\left(\MonU{\bwp{0}{b'}}{\varphi_1}{\tau}, \InsRobVio{\bwp{0}{b'}}{\varphi_2}{\tau}\right)
                \end{array}
                \right) && \text{\footnotesize(by Defs.~\ref{def:classicMonitor} \& \ref{def:quanFineMonitor})}\\
                = &\min\left(
                \begin{array}{l}
                    \max\left(\MonU{\bwp{0}{b}}{\varphi_1}{\tau}, \MonU{\bwp{0}{b}}{\varphi_2}{\tau}\right),  \\
                    \max\left(\InsRobVio{\bwp{0}{b'}}{\varphi_1}{\tau}, \MonU{\bwp{0}{b}}{\varphi_2}{\tau}\right),\\
                    \max\left(\MonU{\bwp{0}{b}}{\varphi_1}{\tau}, \InsRobVio{\bwp{0}{b'}}{\varphi_2}{\tau}\right),\\
                    \max\left(\InsRobVio{\bwp{0}{b'}}{\varphi_1}{\tau}, \InsRobVio{\bwp{0}{b'}}{\varphi_2}{\tau}\right)
                \end{array}
                \right) &&\text{\footnotesize (by global assump.)}\\
                 = &\max\left(
                \begin{array}{l}
                     \min\left(\MonU{\bwp{0}{b}}{\varphi_1}{\tau}, \InsRobVio{\bwp{0}{b'}}{\varphi_1}{\tau}\right),  \\
                      \min\left(\MonU{\bwp{0}{b}}{\varphi_2}{\tau}, \InsRobVio{\bwp{0}{b'}}{\varphi_2}{\tau}\right)
                \end{array}
                \right) && \text{\footnotesize(by def. of min, max)}\\
               =&\max\left(\MonU{\bwp{0}{b'}}{\varphi_1}{\tau}, \MonU{\bwp{0}{b'}}{\varphi_2}{\tau}\right) && \text{\footnotesize (by global assump.)} \\
               =&\MonU{\bwp{0}{b'}}{\varphi_1\lor\varphi_2}{\tau} &&\text{\footnotesize (by Def.~\ref{def:classicMonitor})} \tag*{\qed}
            \end{align*}  
            }
        \end{compactitem}
\end{proof}

Thm.~\ref{theo:reconstruct} shows that the result $\MonU{\bwp{0}{b}}{\varphi}{\tau}$ of \classicMon can be derived from the result of \quanCauseMon by applying $\inf_{t\in[0,b]}\InsRobVio{\bwp{0}{b}}{\varphi}{t}$. For instance, comparing the  results of \quanCauseMon in Fig.~\ref{fig:quanMonitorEg} and the results of \classicMon in Fig.~\ref{fig:runningExample}, we can find that the results in Fig.~\ref{fig:runningExample} can be reconstructed by using the results in Fig.~\ref{fig:quanMonitorEg}.

    

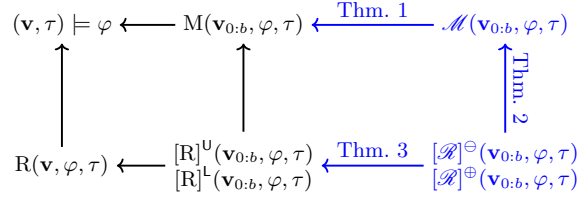
\begin{figure}[h]
    \centering
    \resizebox{0.65\textwidth}{!}{
    \begin{tikzpicture}
    \node (1_1) [] {$(\mathbf{v},\tau) \models \varphi$}; 
    \node (1_2) [right = 2.5em of 1_1] {$\text{M}(\mathbf{v}_{0:b},\varphi,\tau)$}; 
    \node (1_3) [right = 5.5em of 1_2, blue] {$\mathscr{M}(\mathbf{v}_{0:b},\varphi,\tau)$}; 
    \node (2_1) [below = 5.5em of 1_1, anchor=center] {$\mathrm{R}(\mathbf{v},\varphi,\tau)$}; 
    \node (2_2) [below=5.5em of 1_2, align=center,anchor=center]{$\MonU{\bwp{0}{b}}{\varphi}{\tau}$ \\ $\MonL{\bwp{0}{b}}{\varphi}{\tau}$};
    \node (2_3) [below=5.5em of 1_3, align=center,anchor=center, blue] {$[\mathscr{R}]^{\ominus}(\mathbf{v}_{0:b},\varphi,\tau)$ \\ $[\mathscr{R}]^{\oplus}(\mathbf{v}_{0:b},\varphi,\tau)$}; 
    
    \draw[-{To[length=1mm]},thick] (2_1) edge[] (1_1);
    \draw[-{To[length=1mm]},thick] (1_2) edge[] (1_1);
    \draw[-{To[length=1mm]},thick] (2_2) edge[] (2_1);
    \draw[-{To[length=1mm]},thick] (2_2) edge[] (1_2);

    \draw[-{To[length=1mm]},thick] (1_3) edge[above,blue] node[blue] {Thm.~\ref{lem:boolFineMonitor}} (1_2);
    \draw[-{To[length=1mm]},thick] (2_3) edge[right,blue] node[blue] {\rotatebox{-90}{Thm.~\ref{theo:extension}}} (1_3);
    \draw[-{To[length=1mm]},thick] (2_3) edge[above,blue] node[blue] { Thm.~\ref{theo:reconstruct}} (2_2);
    \end{tikzpicture}
    }
    \caption{Refinement among STL monitors}
    \label{fig:refinement}
\end{figure}

\begin{myremark}
\noindent Fig.~\ref{fig:refinement} shows the refinement relations between the six STL monitoring approaches. The left column lists the offline monitoring approaches derived directly from the Boolean and quantitative semantics of STL respectively. The middle column shows the classic online monitoring approaches. Our two causation monitors, namely \boolCauseMon and \quanCauseMon, are given in the column on the right. Given a pair $(A,B)$ of the approaches, $A\leftarrow B$ indicates that the approach $B$ refines the approach $A$, in the sense that $B$ can deliver more information than $A$, and the information delivered by $A$ can be derived from the information delivered by $B$. It is clear that the refinement relation in the figure ensures transitivity. Note that blue arrows are contributed by this paper. As shown by Fig.~\ref{fig:refinement}, the relation between \boolCauseMon and \quanCauseMon is analogous to that between the Boolean and quantitative semantics of STL.
\end{myremark}

\section{Experimental Evaluation}\label{sec:experiment}

We implemented a tool\footnote{Available at \url{https://github.com/choshina/STL-causation-monitor}, and Zenodo~\cite{zhang_zhenya_2023_7923888}.} for our two causation monitors. It is built on the top of Breach~\cite{donze2010breach}, a widely used tool for monitoring and testing of hybrid systems~\cite{ARCHCOMP21Falsification}.
Being consistent with Breach, the monitors target the output signals given by Simulink models, as an additional block. Experiments were executed on a MacOS machine, 1.4 GHz Quad-Core Intel Core-i5, 8 GB RAM, using Breach v1.10.0. 

\subsection{Experiment Setting}\label{sec:experimentSetting}

\myparagraph{Benchmarks} We perform the experiments on the following two benchmarks.

\noindent
\emph{Abstract Fuel Control (AFC)} is a powertrain control system from Toyota~\cite{jin2014powertrain}, which has been widely used as a benchmark in the hybrid system community~\cite{ARCHCOMP20Falsification, ARCHCOMP21Falsification,ARCHCOMP22Falsification}. The system outputs the \emph{air-to-fuel} ratio \af, and requires that the deviation of \af from its reference value \afref should not be too large. Specifically, we consider the following properties from different perspectives:
\begin{compactitem}
    \item $\spec{AFC}{1} \Defeq \Box_{[10, 50]}(\left|\af-\afref\right| < 0.1 )$: the deviation should always be small;
    \item  $\spec{AFC}{2} \Defeq \Box_{[10, 48.5]}\Diamond_{[0,1.5]}\left(\left|\af-\afref\right| < 0.08\right)$: a large deviation should not last for too long time;
    \item $\spec{AFC}{3} \Defeq \Box_{[10,48]}(|\af-\afref | > 0.08 \to \Diamond_{[0,2]}(|\af -\afref| < 0.08))$: whenever the deviation is too large, it should recover to the normal status soon. 
\end{compactitem}

\smallskip
\noindent
\emph{Automatic transmission (AT)} is a widely-used benchmark~\cite{ARCHCOMP20Falsification, ARCHCOMP21Falsification,ARCHCOMP22Falsification}, implementing the transmission controller of an automotive system. 
It outputs the $\gear$, $\speed$ and $\rpm$ of the vehicle, which are required to satisfy this safety requirement:
\begin{compactitem}
    \item $\spec{AT}{1} \Defeq \Box_{[0,27]}(\speed > 50 \to \Diamond_{[1,3]}(\rpm < 3000))$: whenever the \speed is higher than 50, the \rpm should be below 3000 in three time units.
\end{compactitem}

\myparagraph{Baseline and experimental design}
In order to assess our two proposed monitors (the Boolean causation monitor \boolCauseMon in Def.~\ref{def:boolFineMonitor}, and the quantitative causation monitor \quanCauseMon in Def.~\ref{def:quanFineMonitor}), we compare them with two baseline monitors: the classic quantitative robustness monitor \classicMon  (see Def.~\ref{def:classicMonitor}); and the state-of-the-art approach \emph{monitor with reset} \resetMon~\cite{onlineResetTCAD2022}, that, once the signal violates the specification, resets at that point and forgets the previous partial signal.

Given a model and a specification, we generate input signals by randomly sampling in the input space and feed them to the model. The online output signals are given as inputs to the monitors and the monitoring results are collected. We generate 10 input signals for each model and specification. To account for fluctuation of monitoring times in different repetitions\footnote{Note that only the monitoring time changes across different repetitions; monitoring results are instead always the same, as monitoring is deterministic for a given signal.}, for each signal, the experiment has been executed 10 times, and we report average results.

\subsection{Evaluation}\label{sec:evaluation}

\begin{figure}[!tb]
\centering
\begin{subfigure}[b]{0.49\columnwidth}
\centering
\includegraphics[width=0.9\linewidth]{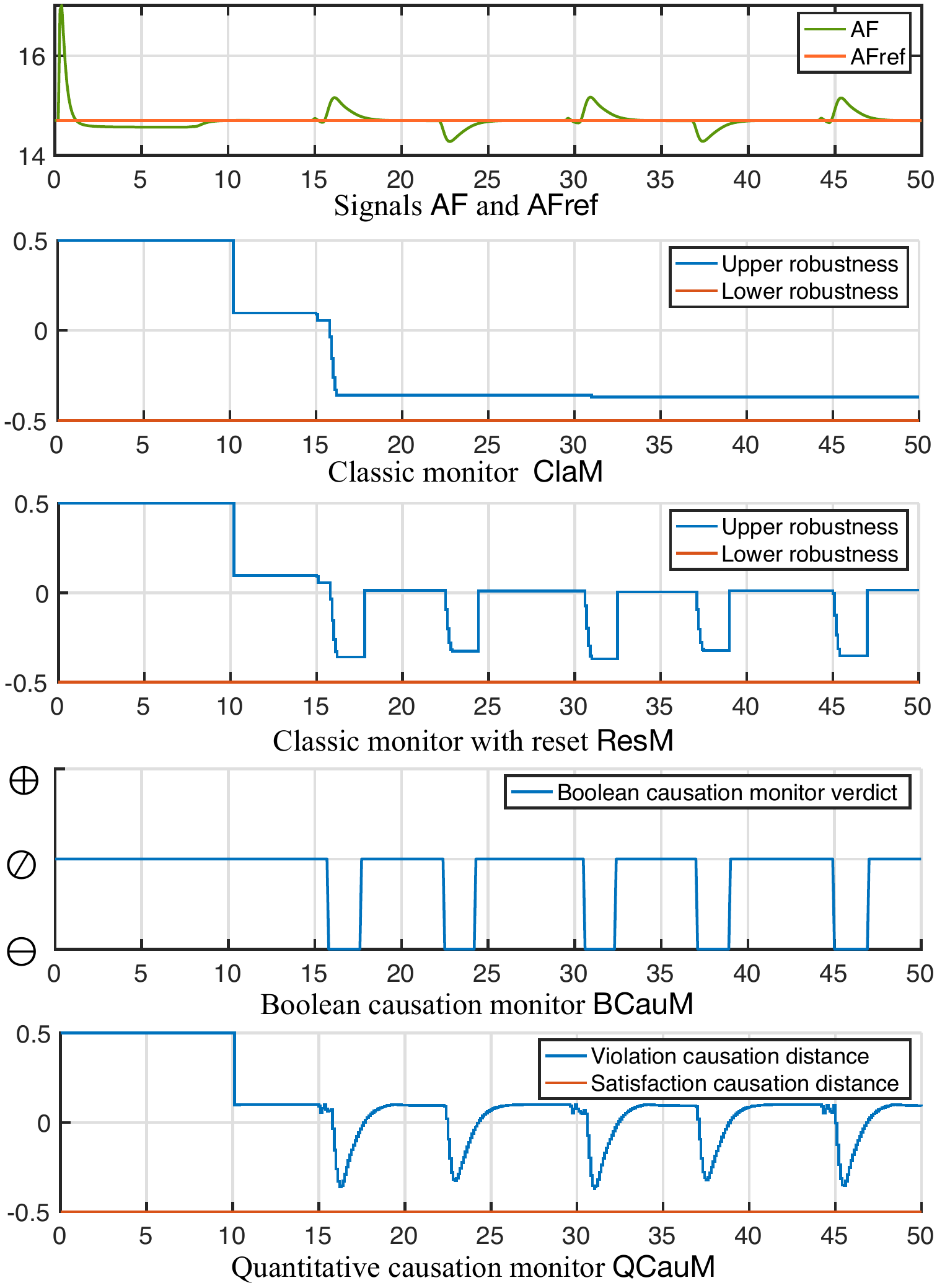}
\caption{Specification \spec{AFC}{1} and signal \sig{4}}
\end{subfigure}
\hfill
\begin{subfigure}[b]{0.49\columnwidth}
\centering
\includegraphics[width=0.9\linewidth]{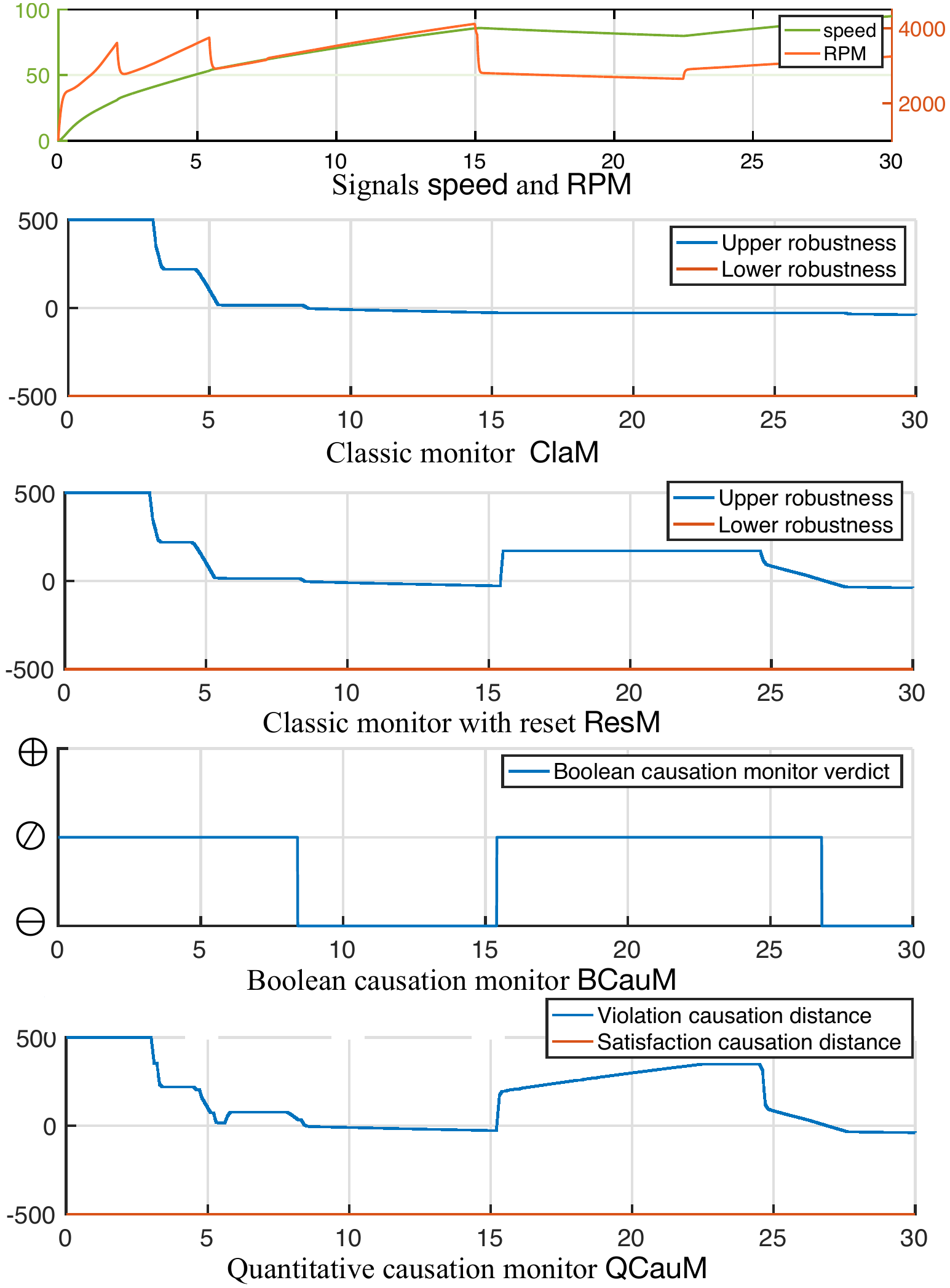}
\caption{Specification \spec{AT}{1} and signal \sig{8}}
\end{subfigure}
\caption{Examples of the information provided by the different monitors}
\label{fig:effect}	
\end{figure}

\myparagraph{Qualitative evaluation}
We here show the type of information provided by the different monitors. As an example, Fig.~\ref{fig:effect} reports, for two specifications of the two models, the system output signal (in the top of the two sub-figures), and the monitoring results of the compared monitors.
We notice that signals of both models (top plots) violate the corresponding specifications in multiple points. Let us consider monitoring results of \spec{AFC}{1}; similar observations apply to \spec{AT}{1}.

When using the \classicMon, only the first violation right after time $15$ is detected (the upper bound of robustness becomes negative); after that, the upper bound remains constant, without reporting that the system recovers from violation at around time $17$, and that the specification is violated again four more times.

Instead, we notice that the monitor with reset \resetMon is able to detect all the violations (as the upper bound becomes greater than $0$ when the violation episode ends), but it does not properly report the margin of robustness; indeed, during the violation episodes, it reports a constant value of around $-0.4$ for the upper bound, but the system violates the specification with different degrees of severity in these intervals; in a similar way, when the specification is satisfied around after time $17$, the upper bound is just above $0$, but actually the system satisfies the specification with different margins. As a consequence, \resetMon provides sharp changes of the robustness upper bound that do not faithfully reflect the system evolution.

We notice that the Boolean causation monitor \boolCauseMon only reports information about the violation episodes, but not on the degree of violation/satisfaction. Instead, the quantitative causation monitor \quanCauseMon is able to provide a very detailed information, not only reporting all the violation episodes, but also properly characterizing the degree with which the specification is violated or satisfied. Indeed, in \quanCauseMon, the violation causation distance smoothly increases from violation to satisfaction, so faithfully reflecting the system evolution.

\myparagraph{Quantitative assessment of monitoring time}
We discuss the computation cost of doing the monitoring.

\begin{table}[!tb]
\caption{Experimental results -- Average (avg.) and standard deviation (stdv.) of monitoring and simulation times (ms)}
\label{table:averageResult}
\resizebox{\textwidth}{!}{%
\begin{tabular}{l rrcrr c rrcrr c rrcrr c rrcrr}
\toprule
& \multicolumn{5}{c}{\classicMon}  & & \multicolumn{5}{c}{\resetMon}  & & \multicolumn{5}{c}{\boolCauseMon}  & & \multicolumn{5}{c}{\quanCauseMon}    \\ \cmidrule{2-6} \cmidrule{8-12} \cmidrule{14-18} \cmidrule{20-24}
     & \multicolumn{2}{c}{monitor} & & \multicolumn{2}{c}{total} & & \multicolumn{2}{c}{monitor} && \multicolumn{2}{c}{total} && \multicolumn{2}{c}{monitor} && \multicolumn{2}{c}{total} && \multicolumn{2}{c}{monitor} && \multicolumn{2}{c}{total} \\ 
& avg. & stdv.  && avg.  & stdv.   && avg.  & stdv.    && avg.  & stdv.         && avg.  & stdv.       && avg.  & stdv.     && avg. & stdv.       && avg.  & stdv.         \\
\midrule
\spec{AFC}{1} & 14.6 & 0.1 && 982.8 & 3.5 && 8.8 & 2.4 && 981.3  & 6.7 && 36.9 & 5.4 && 1009.7 & 16.5 && 15.1 & 0.1 && 981.9 & 4.4 \\
\spec{AFC}{2} & 26.8 & 0.2 && 998.5 & 9.0 && 20.2 & 5.2 && 988.0 & 9.9  && 50.4 & 22.4 && 1023.9 & 25.1 && 27.4 & 0.2 && 999.5 & 8.2 \\
\spec{AFC}{3} & 42.0 & 0.3 && 1016.5 & 8.9 && 45.5 & 4.8 && 1016.9 & 7.5 && 48.4 & 6.2 && 1021.2 & 7.9 && 81.0 & 1.2 && 1060.1 & 5.3  \\
\spec{AT}{1} & 16.7 & 0.2 && 966.0 & 2.6 && 24.0 & 17.0 && 980.4 & 24.2 && 96.1 & 82.6 && 1065.2 & 93.4 && 31.2 & 0.6 && 985.0 & 7.5\\
\bottomrule
\end{tabular}
}
\end{table}
In Table~\ref{table:averageResult}, we observe that, for all the monitors, the \emph{monitor}ing time is much lower than the \emph{total} time (system execution + monitoring). It shows that, for this type of systems, the monitoring overhead is negligible. Still, we compare the execution costs for the different monitors. Table~\ref{table:experiment_result} reports the monitoring times of all the monitors for each specification and each signal.
\begin{table}[!t]
\centering
\caption{Experimental results of the four monitoring approaches -- Monitoring time (ms) -- $\Delta A = \nicefrac{(\quanCauseMon - A)}{A}$}
\label{table:experiment_result}
\begin{subtable}{0.49\textwidth}
\resizebox{\textwidth}{!}{%
\begin{tabular}{lrrrrrrr}
\toprule
\multirow{2}{*}{\textcolor{blue}{$\spec{AFC}{1}$}} & \multirow{2}{*}{\classicMon} & \multirow{2}{*}{\resetMon} & \multirow{2}{*}{\boolCauseMon} & \multirow{2}{*}{\quanCauseMon} & \multicolumn{3}{c}{\quanCauseMon stat. (\%)} \\ \cmidrule{6-8}
  &  &  &  &  & $\Delta$\classicMon & $\Delta$\resetMon & $\Delta$\boolCauseMon  \\
\midrule
\sig{1}                    & 14.5                     & 8.2                    & 37.4                   & 15.2                   & 4.8       & 85.4        & -59.4      \\
\sig{2}                    & 14.5                     & 8.1                    & 39.9                   & 15.0                   & 3.4       & 85.2        & -62.4      \\
\sig{3}                   & 14.8                     & 8.0                    & 38.2                   & 15.0                   & 1.4       & 87.5        & -60.7      \\
\sig{4}                   & 14.7                     & 8.5                    & 38.8                   & 15.3                   & 4.1       & 80.0        & -60.6      \\
\sig{5}                   & 14.6                     & 8.0                    & 37.3                   & 14.9                   & 2.1       & 86.3        & -60.1      \\
\sig{6}                    & 14.6                     & 8.2                    & 37.6                   & 15.1                   & 3.4       & 84.1        & -59.8      \\
\sig{7}                    & 14.6                     & 15.5                   & 21.6                   & 15.0                   & 2.7       & -3.2        & -30.6      \\
\sig{8}                   & 14.7                     & 7.9                    & 39.5                   & 15.0                   & 2.0       & 89.9        & -62.0      \\
\sig{9}                   & 14.6                     & 7.8                    & 39.9                   & 15.1                   & 3.4       & 93.6        & -62.2      \\
\sig{10}                   & 14.5                     & 8.0                    & 38.4                   & 15.1                   & 4.1       & 88.8        & -60.7     \\
\bottomrule
\end{tabular}
}
\end{subtable}
\hfill
\begin{subtable}{0.49\textwidth}
\resizebox{\textwidth}{!}{%
\begin{tabular}{lrrrrrrr}
\toprule
\multirow{2}{*}{\textcolor{blue}{$\spec{AFC}{2}$}} & \multirow{2}{*}{\classicMon} & \multirow{2}{*}{\resetMon} & \multirow{2}{*}{\boolCauseMon} & \multirow{2}{*}{\quanCauseMon} & \multicolumn{3}{c}{\quanCauseMon stat. (\%)} \\ \cmidrule{6-8}
 &  &  &  &  & $\Delta$\classicMon & $\Delta$\resetMon & $\Delta$\boolCauseMon \\
\midrule
\sig{1}                    & 26.8                 & 19.8                 & 45.9                   & 27.4                   & 2.2       & 38.4        & -40.3      \\
\sig{2}                    & 27.1                 & 27.3                 & 27.6                   & 27.8                   & 2.6       & 1.8         & 0.7        \\
\sig{3}                    & 26.6                 & 26.2                 & 30.0                   & 27.5                   & 3.4       & 5.0         & -8.3       \\
\sig{4}                    & 26.6                 & 14.2                 & 107.2                  & 27.0                   & 1.5       & 90.1        & -74.8      \\
\sig{5}                    & 26.7                 & 15.8                 & 50.9                   & 27.3                   & 2.2       & 72.8        & -46.4      \\
\sig{6}                    & 26.6                 & 15.8                 & 56.4                   & 27.2                   & 2.3       & 72.2        & -51.8      \\
\sig{7}                    & 26.8                 & 25.4                 & 33.5                   & 27.5                   & 2.6       & 8.3         & -17.9      \\
\sig{8}                    & 26.9                 & 17.0                 & 51.9                   & 27.4                   & 1.9       & 61.2        & -47.2      \\
\sig{9}                    & 27.1                 & 25.1                 & 50.9                   & 27.6                   & 1.8       & 10.0        & -45.8      \\
\sig{10}                   & 26.7                 & 15.8                 & 50.1                   & 27.3                   & 2.2       & 72.8        & -45.5   \\
\bottomrule
\end{tabular}
}
\end{subtable}
\\
\begin{subtable}{0.49\textwidth}
\resizebox{\textwidth}{!}{%
\begin{tabular}{lrrrrrrr}
\toprule
\multirow{2}{*}{\textcolor{blue}{$\spec{AFC}{3}$}} & \multirow{2}{*}{\classicMon} & \multirow{2}{*}{\resetMon} & \multirow{2}{*}{\boolCauseMon} & \multirow{2}{*}{\quanCauseMon} & \multicolumn{3}{c}{\quanCauseMon stat. (\%)} \\ \cmidrule{6-8}
 &  &  &  &  & $\Delta$\classicMon & $\Delta$\resetMon & $\Delta$\boolCauseMon \\
 \midrule
\sig{1}                    & 42.1                 & 49.2                 & 49.1                   & 81.2                   & 92.9      & 65.0        & 65.4       \\
\sig{2}                   & 42.5                 & 42.2                 & 42.2                   & 82.1                   & 93.2      & 94.5        & 94.5       \\
\sig{3}                    & 41.8                 & 48.8                 & 48.8                   & 81.5                   & 95.0      & 67.0        & 67.0       \\
\sig{4}                    & 42.0                 & 34.9                 & 63.4                   & 78.8                   & 87.6      & 125.8       & 24.3       \\
\sig{5}                    & 41.7                 & 48.9                 & 48.7                   & 79.6                   & 90.9      & 62.8        & 63.4       \\
\sig{6}                    & 41.7                 & 48.5                 & 48.7                   & 79.7                   & 91.1      & 64.3        & 63.7       \\
\sig{7}                    & 42.3                 & 42.7                 & 42.5                   & 81.9                   & 93.6      & 91.8        & 92.7       \\
\sig{8}                    & 42.1                 & 42.2                 & 42.0                   & 81.6                   & 93.8      & 93.4        & 94.3       \\
\sig{9}                    & 42.3                 & 49.1                 & 49.3                   & 82.6                   & 95.3      & 68.2        & 67.5       \\
\sig{10}                   & 41.6                 & 48.6                 & 49.1                   & 80.8                   & 94.2      & 66.3        & 64.6  \\
\bottomrule
\end{tabular}
}
\end{subtable}
\hfill
\begin{subtable}{0.49\textwidth}
\resizebox{\textwidth}{!}{%
\begin{tabular}{lrrrrrrr}
\toprule
\multirow{2}{*}{\textcolor{blue}{$\spec{AT}{1}$}} & \multirow{2}{*}{\classicMon} & \multirow{2}{*}{\resetMon} & \multirow{2}{*}{\boolCauseMon} & \multirow{2}{*}{\quanCauseMon} & \multicolumn{3}{c}{\quanCauseMon stat. (\%)} \\ \cmidrule{6-8}
&  &  &  &  &  $\Delta$\classicMon & $\Delta$\resetMon &  $\Delta$\boolCauseMon \\
\midrule
\sig{1}                    & 16.9                 & 30.7                 & 29.6                   & 32.1                   & 89.9      & 4.6         & 8.4        \\
\sig{2}                    & 16.7                 & 17.4                 & 17.4                   & 31.9                   & 91.0      & 83.3        & 83.3       \\
\sig{3}                    & 16.7                 & 16.8                 & 253.4                  & 31.0                   & 85.6      & 84.5        & -87.8      \\
\sig{4}                   & 16.9                 & 69.7                 & 70.2                   & 31.8                   & 88.2      & -54.4       & -54.7      \\
\sig{5}                    & 16.8                 & 19.6                 & 135.9                  & 31.0                   & 84.5      & 58.2        & -77.2      \\
\sig{6}                    & 16.5                 & 26.5                 & 200.5                  & 30.2                   & 83.0      & 14.0        & -84.9      \\
\sig{7}                    & 16.6                 & 14.6                 & 37.9                   & 31.0                   & 86.7      & 112.3       & -18.2      \\
\sig{8}                    & 16.8                 & 16.4                 & 143.8                  & 31.4                   & 86.9      & 91.5        & -78.2      \\
\sig{9}                   & 16.3                 & 13.9                 & 38.6                   & 31.0                   & 90.2      & 123.0       & -19.7      \\
\sig{10}                   & 16.5                 & 14.2                 & 33.2                   & 30.9                   & 87.3      & 117.6       & -6.9  \\
\bottomrule
\end{tabular}
}
\end{subtable}
\end{table}
Moreover, it reports the percentage difference between the quantitative causation monitor \quanCauseMon (the most informative one) and the other monitors.

We first observe that \resetMon and \boolCauseMon have, for the same specification, high variance of the monitoring times across different signals. \classicMon and \quanCauseMon, instead, provide very consistent monitoring times. This is confirmed by the standard deviation results in Table~\ref{table:averageResult}. The consistent monitoring cost of \quanCauseMon is a good property, as the designers of the monitor can precisely forecast how long the monitoring will take, and design the overall system accordingly.

We observe that \quanCauseMon is negligibly slower than \classicMon for \spec{AFC}{1} and \spec{AFC}{2}, and at most twice slower for the other two specifications. This additional monitoring cost is acceptable, given the additional information provided by \quanCauseMon. Compared to \resetMon, \quanCauseMon is usually slower (at most around the double); also in this case, as \quanCauseMon provides more information than \resetMon, the cost is acceptable.

Compared to the Boolean causation monitor \boolCauseMon, \quanCauseMon is usually faster, as it does not have to collect epochs, which is a costly operation. However, we observe that it is slower in \spec{AFC}{3}, because, in this specification, most of the signals do not violate it (and so also \boolCauseMon does not collect epochs in this case).

To conclude, \quanCauseMon is a monitor able to provide much more information that exiting monitors, with an acceptable overhead in terms of monitoring time.

\section{Related Work}\label{sec:relatedWork}

\myparagraph{Monitoring of STL}
Monitoring can be performed either offline or online. Offline monitoring~\cite{maler2004monitoring,NickovicM07,DonzeFM13} targets complete traces and returns either $\true$ or $\false$. 
In contrast, online monitoring deals with the partial traces, and thus a three-valued semantics was introduced for LTL monitoring~\cite{BauerLS06,bauer2011runtime}, and in further for MTL and STL qualitative online monitoring~\cite{ho2014online,MalerN13}, to handle the situation where neither of the conclusiveness can be made. In usual, the quantitative online monitoring provides a quantitative value or a robust satisfaction interval~\cite{dokhanchi2014line,DokhanchiHF15,jakvsic2015signal,deshmukh2017robust,jakvsic2018quantitative}. Based on them, several tools have been developed, e.g., AMT~\cite{NickovicM07,nivckovic2020amt}, Breach~\cite{donze2010breach}, S-Taliro~\cite{AnnpureddyLFS11}, etc. We refer to the survey~\cite{BartocciDDFMNS18} for comprehensive introduction. Recently, in \cite{QinD20}, Qin and Deshmukh propose clairvoyant monitoring to forecast future signal values and give probabilistic bounds on the specification validity. In~\cite{BalakrishnanRV2021}, an online monitoring is proposed for perception systems with Spatio-temporal Perception Logic~\cite{STPL}.

\myparagraph{Monotonicity issue} However, most of these works do not handle the monotonicity issue stated in this paper. In~\cite{cimatti2019assumption}, Cimatti et al. propose an assumption-based monitoring framework for LTL. It takes the user expertise into account and allows the monitor \emph{resettable}, in the sense that it can restart from any discrete time point. In~\cite{selyunin2017runtime}, a recovery feature is introduced in their online monitor~\cite{jakvsic2015signal}. However, the technique is an application-specific approach, rather than a general framework. In~\cite{onlineResetTCAD2022}, a reset mechanism is proposed for STL online monitor. However, as experimentally evaluated in \S{}\ref{sec:experiment}, it essentially provides a solution for the Boolean semantics and still holds monotonicity between two resetting points.

\myparagraph{Signal diagnostics}
Signal diagnostics~\cite{bartocci2018localizing,ferrere2015trace, nivckovic2020amt} is originally used in an offline manner, for the purpose of fault localization and system debugging. In~\cite{ferrere2015trace}, the authors propose an approach to automatically address the single evaluations (namely, epochs) that account for the satisfaction/violation of an STL specification, for a complete trace. This information can be further used as a reference for detecting the root cause of the bugs in the CPS systems~\cite{nivckovic2020amt, bartocci2021cpsdebug, bartocci2018localizing}. The online version of signal diagnostics, which is the basis of our Boolean causation monitor, is introduced in~\cite{onlineResetTCAD2022}. However, we show in \S{}\ref{sec:experiment} that the monitor based on this technique is often costly, and not able to deliver the quantitative runtime information compared to the quantitative causation monitor. 

\section{Conclusion and Future Work}\label{sec:conclusions}

In this paper, we propose a new way of doing STL monitoring based on causation that is able to provide more information than classic monitoring based on STL robustness. Concretely, we propose two causation monitors, namely \boolCauseMon and \quanCauseMon. In particular, \boolCauseMon intuitively explains the concept of ``causation'' monitoring, and thus paves the path to \quanCauseMon that is more practically valuable.  We further prove the relation between the proposed causation monitors and the classic ones. 

As future work, we plan to improve the efficiency the monitoring, by avoiding some unnecessary computations for some instants. Moreover, we plan to apply it to the monitoring of real-world systems.

%
%
%
\bibliographystyle{splncs04}
\bibliography{biblio}

\newpage

\appendix
\section{Complete Proofs for the Theorems and Lemmas}
\subsection{Proof for Lemma~\ref{lem:STLmono}}\label{sec:monotonicityProof}
\begin{proof}
In this proof, 
we let $b_1$ and $b_2$ be two instants such that $0 \le b_1 < b_2 \le T$, and we prove that it holds that $\MonU{\bwp{0}{b_1}}{\varphi}{\tau} \ge \MonU{\bwp{0}{b_2}}{\varphi}{\tau}$ and $\MonL{\bwp{0}{b_1}}{\varphi}{\tau} \le \MonL{\bwp{0}{b_2}}{\varphi}{\tau}$, by induction on the structure of the STL formula $\varphi$.

First, we prove that Lemma~\ref{lem:STLmono} holds for the atomic propositions. Let $\varphi$ be an atomic proposition $\alpha$. Then it holds the following three cases:
\begin{compactitem}
\item if $b_1< b_2 < \tau$, then we have $\MonU{\bwp{0}{b_1}}{\alpha}{\tau} = \MonU{\bwp{0}{b_2}}{\alpha}{\tau} = \Rmax$ and $\MonL{\bwp{0}{b_1}}{\alpha}{\tau} = \MonL{\bwp{0}{b_2}}{\alpha}{\tau} = \Rmin$; 
\item if $b_1  < \tau < b_2$, then $\MonU{\bwp{0}{b_1}}{\alpha}{\tau} = \Rmax$ and $\MonU{\bwp{0}{b_2}}{\alpha}{\tau} = f(\bw(\tau))\le \MonU{\bwp{0}{b_1}}{\alpha}{\tau}$; similarly,  $\MonL{\bwp{0}{b_1}}{\alpha}{\tau} = \Rmin$ and $\MonL{\bwp{0}{b_2}}{\alpha}{\tau} = f(\bw(\tau)) \ge \MonL{\bwp{0}{b_1}}{\alpha}{\tau}$;
\item if $\tau < b_1 < b_2$, then $\MonU{\bwp{0}{b_1}}{\alpha}{\tau} = \MonU{\bwp{0}{b_2}}{\alpha}{\tau} = f(\bw(\tau))$ and $\MonL{\bwp{0}{b_1}}{\alpha}{\tau} = \MonL{\bwp{0}{b_2}}{\alpha}{\tau} = f(\bw(\tau))$.
\end{compactitem}
Therefore, the lemma holds for atomic propositions.

Second, we assume that the theorem holds for an arbitrary formula $\varphi$, that is, given $\Mon{\bwp{0}{b_1}}{\varphi'}{\tau} = [l_1, u_1]$ and $\Mon{\bwp{0}{b_2}}{\varphi'}{\tau} = [l_2, u_2]$, it holds that $u_2 \le u_1$ and $l_2 \ge l_1$.
We prove that Lemma~\ref{lem:STLmono} also holds for $\varphi'$, which
is constructed by applying the following operators to $\varphi$. 
\begin{compactitem}
\item Case $\varphi' = \neg\varphi$: $\Mon{\bwp{0}{b_1}}{\neg\varphi}{\tau} = -\Mon{\bwp{0}{b_1}}{\varphi}{\tau} = [-u_1, -l_1]$; similarly, $\Mon{\bwp{0}{b_2}}{\neg\varphi}{\tau} = [-u_2, -l_2]$. By the assumption on $\varphi$, it holds that $-u_2\ge -u_1$ and $-l_2 \le -l_1$, so the lemma holds in this case.
\item Case $\varphi' = \varphi_1\land \varphi_2$: we assume that both $\varphi_1$ and $\varphi_2$ satisfy our assumption on $\varphi$. For convenience, we write that $\Mon{\bwp{0}{b_k}}{\varphi_i}{\tau} = \left[l_k^{\varphi_i}, u_k^{\varphi_i}\right]$. Then,
\begin{align*}
    \Mon{\bwp{0}{b_1}}{\varphi_1\land 
    \varphi_2}{\tau}  =& \min\left(\Mon{\bwp{0}{b_1}}{\varphi_1}{\tau}, \Mon{\bwp{0}{b_1}}{\varphi_2}{\tau}\right)\\
    =& \min\left(\left[l_1^{\varphi_1}, u_1^{\varphi_1}\right], \left[l_1^{\varphi_2}, u_1^{\varphi_2}\right]\right) \\
    =& \left[\min\left(l_1^{\varphi_1}, l_1^{\varphi_2}\right), \min\left(u_1^{\varphi_1}, u_1^{\varphi_2}\right)\right]
\end{align*}
similarly, $\Mon{\bwp{0}{b_2}}{\varphi_1\land \varphi_2}{\tau} = \left[\min\left(l_2^{\varphi_1}, l_2^{\varphi_2}\right), \min\left(u_2^{\varphi_1}, u_2^{\varphi_2}\right)\right]$. 
By the assumption that $u_2^{\varphi_1}\le u_1^{\varphi_1}$ and $u_2^{\varphi_2}\le u_1^{\varphi_2}$, it holds that $\min\left(u_2^{\varphi_1}, u_2^{\varphi_2}\right)\le \min\left(u_1^{\varphi_1}, u_1^{\varphi_2}\right)$; and similarly, $\min\left(l_2^{\varphi_1}, l_2^{\varphi_2}\right)\ge \min\left(l_1^{\varphi_1}, l_1^{\varphi_2}\right)$. Therefore, the lemma holds in this case.
\item Case $\varphi' = \Box_I\varphi$: let $\Mon{\bwp{0}{b_1}}{\Box_I\varphi}{\tau} = \displaystyle\inf_{t\in\tau + I}{\left(\Mon{\bwp{0}{b_1}}{\varphi}{t}\right)} = \left[l_1^*, u_1^*\right]$ and $\Mon{\bwp{0}{b_2}}{\Box_I\varphi}{\tau} = \displaystyle\inf_{t\in\tau + I}{\left(\Mon{\bwp{0}{b_2}}{\varphi}{t}\right)} = \Big[l_2^*, u_2^*\Big]$. Moreover, for any $t\in \tau + I$, we write  $\Mon{\bwp{0}{b_1}}{\varphi}{t} = \Big[l_1^t, u_1^t\Big]$, and $\Mon{\bwp{0}{b_2}}{\varphi}{t} = \Big[l_2^t, u_2^t\Big]$. 

By assumption on $\varphi$, it holds that $u_2^t \le u_1^t$ and $l_2^t\ge l_1^t$. Since $u_1^* = \displaystyle\inf_{t\in\tau + I}\left(u_1^t\right)$ and $u_2^* = \displaystyle\inf_{t\in\tau + I}\left(u_2^t\right)$, it holds that $u_2^*\le u_1^*$. Similarly, it holds that $l_2^* \ge l_1^*$. Therefore, the lemma holds in this case.
\end{compactitem}
The proofs for other connectives, such as $\lor$, $\Diamond_I$, $\UntilOp{I}$ follow similar proof patterns, and therefore we skip them. 
\end{proof}

\subsection{Proof for Theorem~\ref{lem:boolFineMonitor}}\label{sec:boolCauseMonProof}

Before proving Thm.~\ref{lem:boolFineMonitor}, we first introduce two lemmas, namely, Lem.~\ref{lem:notEmpty} and Lem.~\ref{lem:firstVio}, about the properties of epochs in Def.~\ref{def:signalDiagnostics}.

\begin{mylemma}\label{lem:notEmpty}
    If a violation epoch $\DiagFSimp{\bwp{0}{b}}{\varphi}{\tau}$ is not empty, it implies that $\MonU{\bwp{0}{b}}{\varphi}{\tau} < 0$; if a satisfaction epoch $\DiagTSimp{\bwp{0}{b}}{\varphi}{\tau}$ is not empty, it implies that $\MonL{\bwp{0}{b}}{\varphi}{\tau} > 0$; if both $\DiagFSimp{\bwp{0}{b}}{\varphi}{\tau}$ and  $\DiagTSimp{\bwp{0}{b}}{\varphi}{\tau}$ are empty, it implies that $\MonU{\bwp{0}{b}}{\varphi}{\tau} > 0$ and $\MonL{\bwp{0}{b}}{\varphi}{\tau} < 0$.
\end{mylemma}

\begin{proof}
    First, it is straightforward to see that the lemma holds if $\varphi$ is an atomic proposition. 

    Then, assuming the lemma holds for an arbitrary formula $\varphi$, we prove that the lemma also holds for $\varphi'$ which is constructed by applying the following operators to $\varphi$.
    \begin{compactitem}
        \item Case $\neg\varphi$: As $\DiagFSimp{\bwp{0}{b}}{\neg\varphi}{\tau}$ is not empty, $\DiagTSimp{\bwp{0}{b}}{\varphi}{\tau}$ is not empty. Then by assumption, it holds that $\MonL{\bwp{0}{b}}{\varphi}{\tau} > 0$, and so $\MonU{\bwp{0}{b}}{\neg\varphi}{\tau} < 0$;
        \item Case $\varphi_1\land\varphi_2$: As $\DiagFSimp{\bwp{0}{b}}{\varphi_1\land\varphi_2}{\tau}$ is not empty, either $\DiagFSimp{\bwp{0}{b}}{\varphi_1}{\tau}$ or $\DiagFSimp{\bwp{0}{b}}{\varphi_2}{\tau}$ is not empty. Say $\DiagFSimp{\bwp{0}{b}}{\varphi_1}{\tau}$ is not empty, then by assumption, $\MonU{\bwp{0}{b}}{\varphi_1}{\tau} < 0$. Then by Def.~\ref{def:classicMonitor}, $\MonU{\bwp{0}{b}}{\varphi_1\land\varphi_2}{\tau} < 0$;
        \item Case $\varphi_1\lor\varphi_2$: By Def.~\ref{def:signalDiagnostics}, if $\DiagFSimp{\bwp{0}{b}}{\varphi_1\lor\varphi_2}{\tau}$ is not empty, it implies that neither $\DiagFSimp{\bwp{0}{b}}{\varphi_1}{\tau}$ nor $\DiagFSimp{\bwp{0}{b}}{\varphi_2}{\tau}$ is empty.  Then by assumption, both $\MonU{\bwp{0}{b}}{\varphi_1}{\tau} < 0$ and $\MonU{\bwp{0}{b}}{\varphi_2}{\tau} < 0$, and therefore by Def.~\ref{def:classicMonitor}, it holds that $\MonU{\bwp{0}{b}}{\varphi_1\lor\varphi_2}{\tau} < 0$. 
        \item Case $\Box_I\varphi$: if $\DiagFSimp{\bwp{0}{b}}{\Box_I\varphi}{\tau}$ is not empty, there exists a $t\in\tau+I$ such that $\DiagFSimp{\bwp{0}{b}}{\varphi}{t}$ is not empty. By assumption, $\MonU{\bwp{0}{b}}{\varphi}{t} < 0$. Then by Def.~\ref{def:classicMonitor}, $\MonU{\bwp{0}{b}}{\Box_I\varphi}{\tau} < 0$;
        \item Case $\Diamond_I\varphi$: By Def.~\ref{def:signalDiagnostics}, if $\DiagFSimp{\bwp{0}{b}}{\Diamond_I\varphi}{\tau}$ is not empty, it implies that none of $\DiagFSimp{\bwp{0}{b}}{\varphi}{t}$, where $t\in\tau+I$, is empty. Then by assumption, $\MonU{\bwp{0}{b}}{\varphi}{t} < 0$ holds for any $t\in\tau+I$, and so by Def.~\ref{def:classicMonitor},  $\MonU{\bwp{0}{b}}{\Diamond_I\varphi}{\tau} < 0$. 
    \end{compactitem}
    The proofs for other two propositions are similar, and so we skip them. \qed 
\end{proof}

\begin{mylemma}\label{lem:firstVio}
    If $\MonU{\bwp{0}{b}}{\varphi}{\tau} < 0$ and, for any $b'\in[0,b]$, $\MonU{\bwp{0}{b'}}{\varphi}{\tau} > 0$, it implies that there exists an atomic proposition $\alpha$ such that $\langle\alpha, b\rangle\in \DiagFSimp{\bwp{0}{b}}{\varphi}{\tau}$; if $\MonL{\bwp{0}{b}}{\varphi}{\tau} > 0$ and, for any $b'\in[0,b]$, $\MonL{\bwp{0}{b'}}{\varphi}{\tau} < 0$, it implies that there exists an atomic proposition $\alpha$ such that $\langle\alpha, b\rangle\in \DiagTSimp{\bwp{0}{b}}{\varphi}{\tau}$.
\end{mylemma}

\begin{proof}
    First, it is straightforward to see that the lemma holds if $\varphi$ is an atomic proposition. 

    Then, we assume the lemma holds for an arbitrary formula $\varphi$, we prove that the lemma also holds for $\varphi'$ which is constructed by applying the following operators to $\varphi$.
    \begin{compactitem}
        \item Case $\neg\varphi$: If $\MonU{\bwp{0}{b}}{\neg\varphi}{\tau} < 0$ and, for any $b'\in[0,b]$, $\MonU{\bwp{0}{b'}}{\neg\varphi}{\tau} > 0$, then by Def.~\ref{def:classicMonitor}, it holds that  $\MonL{\bwp{0}{b}}{\varphi}{\tau} > 0$ and for all $b'\in[0,b]$, $\MonL{\bwp{0}{b'}}{\varphi}{\tau} < 0$. By assumption, there exists an atomic proposition $\alpha$ such that $\langle\alpha, b\rangle\in\DiagTSimp{\bwp{0}{b}}{\varphi}{\tau}$, and so by Def.~\ref{def:signalDiagnostics}, $\langle\alpha, b\rangle\in\DiagFSimp{\bwp{0}{b}}{\neg\varphi}{\tau}$;
        
        \item Case $\varphi_1\land\varphi_2$: If $\MonU{\bwp{0}{b}}{\varphi_1\land\varphi_2}{\tau} < 0$ and, for any $b'\in[0,b]$, $\MonU{\bwp{0}{b'}}{\varphi_1\land\varphi_2}{\tau} > 0$, then either $\MonU{\bwp{0}{b}}{\varphi_1}{\tau} < 0$ or $\MonU{\bwp{0}{b}}{\varphi_2}{\tau} < 0$, and for all $b'\in[0,b]$, both $\MonU{\bwp{0}{b'}}{\varphi_1}{\tau} > 0$ and $\MonU{\bwp{0}{b'}}{\varphi_2}{\tau} > 0$. Let $\MonU{\bwp{0}{b}}{\varphi_1}{\tau} < 0$. By assumption, there exists an atomic proposition $\alpha$ such that $\langle\alpha, b\rangle\in \DiagFSimp{\bwp{0}{b}}{\varphi_1}{\tau}$. By Def.~\ref{def:signalDiagnostics}, $\DiagFSimp{\bwp{0}{b}}{\varphi_1\land\varphi_2}{\tau} \supseteq \DiagFSimp{\bwp{0}{b}}{\varphi_1}{\tau}$, so $\langle\alpha, b\rangle\in \DiagFSimp{\bwp{0}{b}}{\varphi_1\land\varphi_2}{\tau}$;
        
        \item 
        Case $\varphi_1\lor\varphi_2$: If $\MonU{\bwp{0}{b}}{\varphi_1\lor\varphi_2}{\tau} < 0$ and, for any $b'\in[0,b]$, $\MonU{\bwp{0}{b'}}{\varphi_1\lor\varphi_2}{\tau} > 0$, then both $\MonU{\bwp{0}{b}}{\varphi_1}{\tau} < 0$ and $\MonU{\bwp{0}{b}}{\varphi_2}{\tau} < 0$, and either $\MonU{\bwp{0}{b'}}{\varphi_1}{\tau} > 0$ for all $b'\in[0,b]$, or $\MonU{\bwp{0}{b'}}{\varphi_2}{\tau} > 0$ for all $b'\in[0,b]$. Let $\MonU{\bwp{0}{b'}}{\varphi_1}{\tau} > 0$ for all $b'\in[0,b]$. By assumption, there exists an atomic proposition $\alpha$ such that $\langle\alpha, b\rangle\in \DiagFSimp{\bwp{0}{b}}{\varphi_1}{\tau}$. By Def.~\ref{def:signalDiagnostics}, $\DiagFSimp{\bwp{0}{b}}{\varphi_1\lor\varphi_2}{\tau} \supseteq \DiagFSimp{\bwp{0}{b}}{\varphi_1}{\tau}$, so $\langle\alpha, b\rangle\in \DiagFSimp{\bwp{0}{b}}{\varphi_1\lor\varphi_2}{\tau}$;
        
        \item Case $\Box_I\varphi$: If $\MonU{\bwp{0}{b}}{\Box_I\varphi}{\tau} < 0$ and, for any $b'\in[0,b]$, $\MonU{\bwp{0}{b'}}{\Box_I\varphi}{\tau} > 0$, then there exists a $t\in\tau+I$ such that  $\MonU{\bwp{0}{b}}{\varphi}{t} < 0$, and for all $t\in\tau+I$, $\MonU{\bwp{0}{b'}}{\varphi}{t} > 0$ for all $b'\in[0,b]$. By assumption, there exists an atomic proposition $\alpha$ such that $\langle\alpha, b\rangle\in \DiagFSimp{\bwp{0}{b}}{\varphi}{t}$. By Def.~\ref{def:signalDiagnostics} $\DiagFSimp{\bwp{0}{b}}{\Box_I\varphi}{\tau} \supseteq \DiagFSimp{\bwp{0}{b}}{\varphi}{t}$, so $\langle\alpha, b\rangle\in \DiagFSimp{\bwp{0}{b}}{\Box_I\varphi}{\tau}$;

        \item Case $\Diamond_I\varphi$: If $\MonU{\bwp{0}{b}}{\Diamond_I\varphi}{\tau} < 0$ and, for any $b'\in[0,b]$, $\MonU{\bwp{0}{b'}}{\Diamond_I\varphi}{\tau} > 0$, then for all $t\in\tau+I$,  $\MonU{\bwp{0}{b}}{\varphi}{t} < 0$, and there exists a $t\in\tau+I$ such that $\MonU{\bwp{0}{b'}}{\varphi}{t} > 0$ for all $b'\in[0,b]$. By assumption, there exists an atomic proposition $\alpha$ such that $\langle\alpha, b\rangle\in \DiagFSimp{\bwp{0}{b}}{\varphi}{t}$. By Def.~\ref{def:signalDiagnostics} $\DiagFSimp{\bwp{0}{b}}{\Diamond_I\varphi}{\tau} \supseteq \DiagFSimp{\bwp{0}{b}}{\varphi}{t}$, so $\langle\alpha, b\rangle\in \DiagFSimp{\bwp{0}{b}}{\Diamond_I\varphi}{\tau}$.
    \end{compactitem}
    The proof for the case of satisfaction epoch follows the similar pattern, and so we skip it. \qed
\end{proof}

\subsubsection{Proof for Theorem~\ref{lem:boolFineMonitor}}
Now we prove Thm.~\ref{lem:boolFineMonitor}, based on  Lem.~\ref{lem:notEmpty} and Lem~\ref{lem:firstVio}.
\begin{proof}
We first prove $\CMon{\bwp{0}{b}}{\varphi}{\tau} = \bot$ \textit{iff.} $\bigvee_{t\in[0,b]}\left(\CDiag{\bwp{0}{t}}{\varphi}{\tau} = \negCause\right)$.
\begin{compactitem}
    \item from left to right: if $\CMon{\bwp{0}{b}}{\varphi}{\tau} = \bot$, by the monotonicity of the online monitor, there exists a $b'\in[0,b]$ such that for all $b''\in[0,b')$, $\CMon{\bwp{0}{b''}}{\varphi}{\tau} =~?$ and $\CMon{\bwp{0}{b'}}{\varphi}{\tau} = \bot$. By Lem.~\ref{lem:firstVio}, it holds that there exists an $\alpha$ such that $\langle\alpha, b'\rangle\in\DiagFSimp{\bwp{0}{b'}}{\varphi}{\tau}$. By Def.~\ref{def:boolFineMonitor},  it holds that $\CDiag{\bwp{0}{b'}}{\varphi}{\tau} = \negCause$, and so $\bigvee_{t\in[0,b]}\left(\CDiag{\bwp{0}{b}}{\varphi}{\tau} = \negCause\right)$.
    \item from right to left: if $\bigvee_{t\in[0,b]}\left(\CDiag{\bwp{0}{b}}{\varphi}{\tau} = \negCause\right)$, there exists a $b'\in[0,b]$ such that $\CDiag{\bwp{0}{b'}}{\varphi}{\tau} = \negCause$, and by Def.~\ref{def:boolFineMonitor}, $\DiagFSimp{\bwp{0}{b'}}{\varphi}{\tau}$ is not empty. By Lem.~\ref{lem:notEmpty}, $\MonU{\bwp{0}{b'}}{\varphi}{\tau} < 0$, then by the monotonicity of the online monitor, $\MonU{\bwp{0}{b}}{\varphi}{\tau} < 0$, and so $\CMon{\bwp{0}{b}}{\varphi}{\tau} = \bot$.
\end{compactitem}
The proofs for the other two propositions are similar, and so we skip them. \qed
\end{proof}


\subsection{Proof for Theorem~\ref{theo:extension}}\label{sec:theo1proof}
\begin{proof}
The proof is generally based on mathematical induction. First, by Def.~\ref{def:quanFineMonitor} and Def.~\ref{def:signalDiagnostics}, it is straightforward  that Thm.~\ref{theo:extension} holds for the atomic propositions.

Then, assuming that Thm.~\ref{theo:extension} holds 
for an arbitrary formula $\varphi$, that is, $\InsRobVio{\bwp{0}{b}}{\varphi}{\tau} < 0$ implies $\CDiag{\bwp{0}{b}}{\varphi}{\tau} = \negCause$, $\InsRobSat{\bwp{0}{b}}{\varphi}{\tau} > 0$ implies $\CDiag{\bwp{0}{b}}{\varphi}{\tau} = \posCause$, and if $\InsRobVio{\bwp{0}{b}}{\varphi}{\tau} > 0$ and $\InsRobSat{\bwp{0}{b}}{\varphi}{\tau} < 0$, it holds that $\CDiag{\bwp{0}{b}}{\varphi}{\tau} = \nothing$, we prove that Thm.~\ref{theo:extension} also holds for $\varphi'$ which is a composite formula of $\varphi$.

We show the proof for that $\InsRobVio{\bwp{0}{b}}{\varphi'}{\tau} < 0$ implies $\CDiag{\bwp{0}{b}}{\varphi'}{\tau} = \negCause$, under the following cases:
\begin{compactitem}
    \item Case $\varphi' = \neg\varphi$: 
    \begin{align*}
        &\InsRobVio{\bwp{0}{b}}{\neg\varphi}{\tau} < 0\\
        \Rightarrow &\InsRobSat{\bwp{0}{b}}{\varphi}{\tau} > 0 && \mycomment{by Def.~\ref{def:quanFineMonitor}} \\
        \Rightarrow & \CDiag{\bwp{0}{b}}{\varphi}{\tau} = \posCause && \mycomment{by assump.}\\
        \Rightarrow & \exists\alpha.\; \langle\alpha, b\rangle\in \DiagTSimp{\bwp{0}{b}}{\varphi}{\tau} && \mycomment{by Def.~\ref{def:boolFineMonitor}} \\
        \Rightarrow & \exists\alpha.\;  \langle\alpha, b\rangle\in \DiagFSimp{\bwp{0}{b}}{\neg\varphi}{\tau} &&\mycomment{by Def.~\ref{def:signalDiagnostics}} \\
        \Rightarrow & \CDiag{\bwp{0}{b}}{\neg\varphi}{\tau} = \negCause &&\mycomment{by Def.~\ref{def:boolFineMonitor}}
    \end{align*}
    
    \item Case $\varphi' = \varphi_1\land\varphi_2$: 
    \begin{align*}
        &\InsRobVio{\bwp{0}{b}}{\varphi_1\land\varphi_2}{\tau} < 0 \\
        \Rightarrow & \InsRobVio{\bwp{0}{b}}{\varphi_1}{\tau}<0 && \mycomment{by Def.~\ref{def:quanFineMonitor} and w.l.o.g.} \\
        \Rightarrow & \CDiag{\bwp{0}{b}}{\varphi_1}{\tau} = \negCause &&\mycomment{by assump.} \\
        \Rightarrow & \exists\alpha.\; \langle\alpha, b\rangle\in \DiagFSimp{\bwp{0}{b}}{\varphi_1}{\tau}  && \mycomment{by Def.~\ref{def:boolFineMonitor}} \\
        \Rightarrow & \MonU{\bwp{0}{b}}{\varphi_1}{\tau} < 0 && \mycomment{by Thm.~\ref{lem:boolFineMonitor}}\\
        \Rightarrow & \MonU{\bwp{0}{b}}{\varphi_1\land \varphi_2}{\tau} < 0 &&\mycomment{by Def.~\ref{def:classicMonitor}} \\
        \Rightarrow & \DiagFSimp{\bwp{0}{b}}{\varphi_1\land\varphi_2}{\tau} \supseteq \DiagFSimp{\bwp{0}{b}}{\varphi_1}{\tau} && \mycomment{by Def.~\ref{def:signalDiagnostics}}\\
        \Rightarrow & \exists\alpha.\; \langle\alpha, b\rangle\in \DiagFSimp{\bwp{0}{b}}{\varphi_1\land\varphi_2}{\tau} &&\mycomment{by def. of $\supseteq$} \\
        \Rightarrow & \CDiag{\bwp{0}{b}}{\varphi_1\land\varphi_2}{\tau} = \negCause && \mycomment{by Def.~\ref{def:boolFineMonitor}}
    \end{align*}
    
    \item Case $\varphi' = \varphi_1\lor \varphi_2$: 
    \begin{align*}
    &\InsRobVio{\bwp{0}{b}}{\varphi_1\lor\varphi_2}{\tau} < 0 \\
    \Rightarrow & \max\left(\InsRobVio{\bwp{0}{b}}{\varphi_1}{\tau}, \MonU{\bwp{0}{b}}{\varphi_2}{\tau}\right)< 0  &&\mycomment{by Def.~\ref{def:quanFineMonitor} and w.l.o.g.}\\
    \Rightarrow & \bigwedge\left(
    \begin{array}{l}
        \InsRobVio{\bwp{0}{b}}{\varphi_1}{\tau} <0   \\
         \MonU{\bwp{0}{b}}{\varphi_2}{\tau} < 0 
    \end{array}
    \right) && \mycomment{by def. of max}\\
    \Rightarrow & \CDiag{\bwp{0}{b}}{\varphi_1}{\tau} = \negCause && \mycomment{by assumption}\\
    \Rightarrow & \DiagFSimp{\bwp{0}{b}}{\varphi_1\lor\varphi_2}{\tau} \supseteq \DiagFSimp{\bwp{0}{b}}{\varphi_1}{\tau} && \mycomment{by Def.~\ref{def:signalDiagnostics} and Thm.~\ref{lem:boolFineMonitor}}\\
    \Rightarrow & \exists \alpha.\; \langle \alpha, b\rangle\in \DiagFSimp{\bwp{0}{b}}{\varphi_1\lor\varphi_2}{\tau} &&\mycomment{by def. of $\supseteq$} \\
    \Rightarrow & \CDiag{\bwp{0}{b}}{\varphi_1\lor\varphi_2}{\tau} = \negCause &&\mycomment{by Def.~\ref{def:boolFineMonitor}}
\end{align*}

    \item Case $\varphi' = \Box_I\varphi$: 
    \begin{align*}
        &\InsRobVio{\bwp{0}{b}}{\Box_I\varphi}{\tau} < 0 \\
        \Rightarrow & \inf_{t\in\tau+I}\InsRobVio{\bwp{0}{b}}{\varphi}{t} < 0 && \mycomment{by Def.~\ref{def:quanFineMonitor}} \\
        \Rightarrow & \InsRobVio{\bwp{0}{b}}{\varphi}{t'} < 0 \text{ s.t. } t' \in\tau+I && \mycomment{by def. of inf} \\
        \Rightarrow & \CDiag{\bwp{0}{b}}{\varphi}{t'} = \negCause  && \mycomment{by assump.}\\
        \Rightarrow & \exists\alpha.\; \langle\alpha, b\rangle\in \DiagFSimp{\bwp{0}{b}}{\varphi}{t'} && \mycomment{by Def.~\ref{def:boolFineMonitor}}\\
        \Rightarrow & \MonU{\bwp{0}{b}}{\varphi}{t'} < 0 && \mycomment{by Thm.~\ref{lem:boolFineMonitor}}\\
        \Rightarrow & \MonU{\bwp{0}{b}}{\Box_I\varphi}{\tau} < 0 &&\mycomment{by Def.~\ref{def:classicMonitor}}\\
        \Rightarrow & \DiagFSimp{\bwp{0}{b}}{\Box_I\varphi}{\tau} \supseteq \DiagFSimp{\bwp{0}{b}}{\varphi}{t'}  && \mycomment{by Def.~\ref{def:signalDiagnostics}}\\
        \Rightarrow & \exists\alpha.\; \langle\alpha, b\rangle\in \DiagFSimp{\bwp{0}{b}}{\Box_I\varphi}{\tau} &&\mycomment{by def. of $\subseteq$}\\
        \Rightarrow & \CDiag{\bwp{0}{b}}{\Box_I\varphi}{\tau} = \negCause &&\mycomment{by Def.~\ref{def:boolFineMonitor}}
    \end{align*}
    \item Case $\varphi' = \Diamond_I\varphi$: 
    \begin{align*}
        &\InsRobVio{\bwp{0}{b}}{\Diamond_I\varphi}{\tau} < 0 \\
        \Rightarrow & \max\left(\inf_{t\in\tau+I}
        \InsRobVio{\bwp{0}{b}}{\varphi}{t}, \MonU{\bwp{0}{b}}{\Diamond_I\varphi}{\tau}
        \right) < 0 && \mycomment{by Def.~\ref{def:quanFineMonitor}} \\
        \Rightarrow & \bigwedge\left(
        \begin{array}{l}
            \displaystyle\inf_{t\in\tau+I}
        \InsRobVio{\bwp{0}{b}}{\varphi}{t} < 0   \\
             \MonU{\bwp{0}{b}}{\Diamond_I\varphi}{\tau} < 0 
        \end{array}
        \right) &&\mycomment{by def. of max} \\
        \Rightarrow & \InsRobVio{\bwp{0}{b}}{\varphi}{t'} < 0\text{ s.t. } t'\in\tau+I && \mycomment{by def. of inf}\\
        \Rightarrow & \CDiag{\bwp{0}{b}}{\varphi}{t'} = \negCause &&\mycomment{by assump.}\\
        \Rightarrow & \exists\alpha.\; \langle\alpha, b\rangle\in \DiagFSimp{\bwp{0}{b}}{\varphi}{t'} &&\mycomment{by Def.~\ref{def:boolFineMonitor}}\\
        \Rightarrow & \DiagFSimp{\bwp{0}{b}}{\Diamond_I\varphi}{\tau} \supseteq \DiagFSimp{\bwp{0}{b}}{\varphi}{t'} && \mycomment{by Def.~\ref{def:signalDiagnostics}}\\
        \Rightarrow & \exists\alpha.\;\langle\alpha, b\rangle\in \DiagFSimp{\bwp{0}{b}}{\Diamond_I\varphi}{\tau} &&\mycomment{by def. of $\supseteq$} \\
        \Rightarrow & \CDiag{\bwp{0}{b}}{\Diamond_I\varphi}{\tau}  = \negCause &&\mycomment{by Def.~\ref{def:boolFineMonitor}}
    \end{align*}

    \item Case $\varphi' = \varphi_1\UntilOp{I}\varphi_2$: 
    \begin{align*}
        &\InsRobVio{\bwp{0}{b}}{\varphi_1\UntilOp{I}\varphi_2}{\tau} < 0 \\
        \Rightarrow &\max\left(
        \begin{array}{l}
             \min\left(
             \begin{array}{l}
                  \displaystyle\inf_{t''\in[\tau, t')}\InsRobVio{\bwp{0}{b}}{\varphi_1}{t''}  \\
                   \InsRobVio{\bwp{0}{b}}{\varphi_2}{t'}
             \end{array}
             \right)  \\
             \MonU{\bwp{0}{b}}{\varphi_1\UntilOp{I}\varphi_2}{\tau} 
        \end{array}
        \right) < 0 \text{ s.t. } t'\in\tau+I && \mycomment{by Def.~\ref{def:quanFineMonitor}} \\
        \Rightarrow & \bigwedge\left(
        \begin{array}{l}
           \bigvee\left(
             \begin{array}{l}
                  \displaystyle\inf_{t''\in[\tau, t')}\InsRobVio{\bwp{0}{b}}{\varphi_1}{t''} <0 \\
                   \InsRobVio{\bwp{0}{b}}{\varphi_2}{t'}<0
             \end{array}
             \right)     \\
              \MonU{\bwp{0}{b}}{\varphi_1\UntilOp{I}\varphi_2}{\tau} < 0
        \end{array}
        \right) && \hspace{-1.5cm}\mycomment{by def. of $\min$, $\max$}
    \end{align*}
    Now we have $\MonU{\bwp{0}{b}}{\varphi_1\UntilOp{I}\varphi_2}{\tau} < 0$, and we respectively discuss the following two cases:
    \begin{compactitem}
            \item 
        if $\InsRobVio{\bwp{0}{b}}{\varphi_2}{t'} < 0$:
        \begin{align*}
            &\InsRobVio{\bwp{0}{b}}{\varphi_2}{t'} < 0 \\
        \Rightarrow & \CDiag{\bwp{0}{b}}{\varphi_2}{t'} = \negCause &&\mycomment{by assump.}\\
        \Rightarrow & \exists\alpha.\; \langle\alpha, b\rangle\in \DiagFSimp{\bwp{0}{b}}{\varphi_2}{t'} &&\mycomment{by Def.~\ref{def:boolFineMonitor}}\\
        \Rightarrow & \DiagFSimp{\bwp{0}{b}}{\varphi_1\UntilOp{I}\varphi_2}{\tau} \supseteq \DiagFSimp{\bwp{0}{b}}{\varphi_2}{t'} && \mycomment{by Def.~\ref{def:signalDiagnostics}}\\
        \Rightarrow & \exists\alpha.\;\langle\alpha, b\rangle\in \DiagFSimp{\bwp{0}{b}}{\varphi_1\UntilOp{I}\varphi_2}{\tau} &&\mycomment{by def. of $\supseteq$} \\
        \Rightarrow & \CDiag{\bwp{0}{b}}{\varphi_1\UntilOp{I}\varphi_2}{\tau}  = \negCause &&\mycomment{by Def.~\ref{def:boolFineMonitor}}
        \end{align*}
        
        \item if $\inf_{t''\in[\tau, t')}\InsRobVio{\bwp{0}{b}}{\varphi_1}{t''}< 0$:
        \begin{align*}
            &\inf_{t''\in[\tau, t')}\InsRobVio{\bwp{0}{b}}{\varphi_1}{t''}< 0 \\
            \Rightarrow & \InsRobVio{\bwp{0}{b}}{\varphi_1}{t^*}< 0 \text{ s.t. } t^*\in[\tau, t') && \mycomment{by def. of inf}\\
            \Rightarrow & \CDiag{\bwp{0}{b}}{\varphi_1}{t^*} = \negCause &&\mycomment{by assump.}\\
        \Rightarrow & \exists\alpha.\; \langle\alpha, b\rangle\in \DiagFSimp{\bwp{0}{b}}{\varphi_1}{t^*} &&\mycomment{by Def.~\ref{def:boolFineMonitor}}\\
        \Rightarrow & \DiagFSimp{\bwp{0}{b}}{\varphi_1\UntilOp{I}\varphi_2}{\tau} \supseteq \DiagFSimp{\bwp{0}{b}}{\varphi_1}{t^*} && \mycomment{by Def.~\ref{def:signalDiagnostics}}\\
        \Rightarrow & \exists\alpha.\;\langle\alpha, b\rangle\in \DiagFSimp{\bwp{0}{b}}{\varphi_1\UntilOp{I}\varphi_2}{\tau} &&\mycomment{by def. of $\supseteq$} \\
        \Rightarrow & \CDiag{\bwp{0}{b}}{\varphi_1\UntilOp{I}\varphi_2}{\tau}  = \negCause &&\mycomment{by Def.~\ref{def:boolFineMonitor}}
        \end{align*}
    
    \end{compactitem}
\end{compactitem}
The proofs for the cases when $\CDiag{\bwp{0}{b}}{\varphi}{\tau} = \posCause$ or  when $\CDiag{\bwp{0}{b}}{\varphi}{\tau} = \nothing$ follow the similar patterns, and so we omit them.
\qed
\end{proof}

\subsection{Proof for Theorem~\ref{theo:reconstruct}}\label{sec:theo2proof}

\begin{proof}
The proof is generally based on mathematical induction.
First, by Def.~\ref{def:quanFineMonitor} and Def.~\ref{def:classicMonitor}, it is straightforward that Thm.~\ref{theo:reconstruct} holds for the atomic propositions.

Then, we make the global assumption that Thm.~\ref{theo:reconstruct} holds for an arbitrary formula $\varphi$, that is, the two cases $\inf_{t\in[0,b]}\InsRobVio{\bwp{0}{t}}{\varphi}{\tau} = \MonU{\bwp{0}{b}}{\varphi}{\tau}$, and $\sup_{t\in[0,b]}\InsRobSat{\bwp{0}{t}}{\varphi}{\tau} = \MonL{\bwp{0}{b}}{\varphi}{\tau}$ holds. Based on that, we prove that Thm.~\ref{theo:reconstruct} also holds for $\varphi'$ which is a composite STL formula of $\varphi$. 

First, we prove $\inf_{t\in[0,b]}\InsRobVio{\bwp{0}{t}}{\varphi'}{\tau} = \MonU{\bwp{0}{b}}{\varphi'}{\tau}$ as follows.  
    \begin{compactitem}
        \item Case $\varphi' = \neg\varphi$: 
        By Def.~\ref{def:quanFineMonitor}, Def.~\ref{def:classicMonitor} and the assumption, it holds that:
        \begin{align*}
            &\inf_{t\in[0,b]}\InsRobVio{\bwp{0}{t}}{\neg\varphi}{\tau}\\
            =& -\sup_{t\in[0,b]}\InsRobSat{\bwp{0}{t}}{\varphi}{\tau} && \text{\footnotesize (by Def.~\ref{def:quanFineMonitor})}\\
            =& -\MonL{\bwp{0}{b}}{\varphi}{\tau}
            = \MonU{\bwp{0}{b}}{\neg\varphi}{\tau} && \text{\footnotesize (by assump. and Def.~\ref{def:classicMonitor})}
        \end{align*}
        
        \item Case $\varphi' = \varphi_1\land\varphi_2$: By Def.~\ref{def:quanFineMonitor}, Def.~\ref{def:classicMonitor} and the assumption, it holds that:
        \begin{align*}
            &\inf_{t\in[0, b]}\InsRobVio{\bwp{0}{t}}{\varphi_1\land\varphi_2}{\tau} \\
            =& \inf_{t\in[0,b]}\left(\min\left(\InsRobVio{\bwp{0}{t}}{\varphi_1}{\tau}, \InsRobVio{\bwp{0}{t}}{\varphi_2}{\tau}\right)\right) &&\text{\footnotesize (by Def.~\ref{def:quanFineMonitor})}\\
            =&\min\left(
            \inf_{t\in[0,b]}\InsRobVio{\bwp{0}{t}}{\varphi_1}{\tau}, \inf_{t\in[0,b]}\InsRobVio{\bwp{0}{t}}{\varphi_2}{\tau}
            \right) &&\text{\footnotesize (by def. of min, inf)}\\
            =&\min\left(\MonU{\bwp{0}{b}}{\varphi_1}{\tau}, \MonU{\bwp{0}{b}}{\varphi_2}{\tau}\right) && \text{\footnotesize (by assump.)}\\
            =&\MonU{\bwp{0}{b}}{\varphi_1\land\varphi_2}{\tau} &&\text{\footnotesize (by Def.~\ref{def:classicMonitor})}
        \end{align*}   

        \item Case $\varphi' = \varphi_1\lor\varphi_2$: 
        We prove Thm.~\ref{theo:reconstruct} holds for this case by induction on the length $b$ of the partial signal $\bwp{0}{b}$.  
        \begin{compactitem}
            \item First, if $b = \tau$, by Def.~\ref{def:quanFineMonitor} and by the assumption, it holds that:
            \begin{align*}
                &\inf_{t\in[0,b]}\InsRobVio{\bwp{0}{t}}{\varphi_1\lor\varphi_2}{\tau} \\
                =& \InsRobVio{\bwp{0}{\tau}}{\varphi_1\lor\varphi_2}{\tau} && \text{\footnotesize (by Def.~\ref{def:quanFineMonitor})}\\
                =& \max\left(\MonU{\bwp{0}{\tau}}{\varphi_1}{\tau}, \MonU{\bwp{0}{\tau}}{\varphi_2}{\tau}\right) && \text{\footnotesize (by assump. and Def.~\ref{def:quanFineMonitor})}\\
                =& \MonU{\bwp{0}{\tau}}{\varphi_1\lor\varphi_2}{\tau} &&\text{\footnotesize (by Def.~\ref{def:classicMonitor})}
            \end{align*}
            \item Then, we make a local assumption that, given an arbitrary $b$, it holds that $\inf_{t\in[0,b]}\InsRobVio{\bwp{0}{t}}{\varphi_1\lor\varphi_2}{\tau} = \MonU{\bwp{0}{b}}{\varphi_1\lor\varphi_2}{\tau}$. We prove that, for $b'$ which is the next sampling point to $b$, it holds that,
            \begin{align*}
& \inf_{t\in[0,b']}\InsRobVio{\bwp{0}{t}}{\varphi_1\lor\varphi_2}{\tau} \\  = &\min\left(\MonU{\bwp{0}{b}}{\varphi_1\lor\varphi_2}{\tau}, \InsRobVio{\bwp{0}{b'}}{\varphi_1\lor\varphi_2}{\tau}\right)  & &\text{\footnotesize (by local assump.)} \\
                = &\min\left(
                \begin{array}{l}
                     \max\left(\MonU{\bwp{0}{b}}{\varphi_1}{\tau}, \MonU{\bwp{0}{b}}{\varphi_2}{\tau}\right),  \\
                     \max\left(\InsRobVio{\bwp{0}{b'}}{\varphi_1}{\tau}, \MonU{\bwp{0}{b'}}{\varphi_2}{\tau}\right), \\
                     \max\left(\MonU{\bwp{0}{b'}}{\varphi_1}{\tau}, \InsRobVio{\bwp{0}{b'}}{\varphi_2}{\tau}\right)
                \end{array}
                \right) && \text{\footnotesize(by Def.~\ref{def:classicMonitor} \& \ref{def:quanFineMonitor})}\\
                = &\min\left(
                \begin{array}{l}
                    \max\left(\MonU{\bwp{0}{b}}{\varphi_1}{\tau}, \MonU{\bwp{0}{b}}{\varphi_2}{\tau}\right),  \\
                    \max\left(\InsRobVio{\bwp{0}{b'}}{\varphi_1}{\tau}, \MonU{\bwp{0}{b}}{\varphi_2}{\tau}\right),\\
                    \max\left(\InsRobVio{\bwp{0}{b'}}{\varphi_1}{\tau}, \InsRobVio{\bwp{0}{b'}}{\varphi_2}{\tau}\right),\\
                    \max\left(\MonU{\bwp{0}{b}}{\varphi_1}{\tau}, \InsRobVio{\bwp{0}{b'}}{\varphi_2}{\tau}\right)
                \end{array}
                \right) &&\text{\footnotesize (by global assump.)}\\
                 = &\max\left(
                \begin{array}{l}
                     \min\left(\MonU{\bwp{0}{b}}{\varphi_1}{\tau}, \InsRobVio{\bwp{0}{b'}}{\varphi_1}{\tau}\right),  \\
                      \min\left(\MonU{\bwp{0}{b}}{\varphi_2}{\tau}, \InsRobVio{\bwp{0}{b'}}{\varphi_2}{\tau}\right)
                \end{array}
                \right) && \text{\footnotesize(def. of min, max)}\\
               =&\max\left(\MonU{\bwp{0}{b'}}{\varphi_1}{\tau}, \MonU{\bwp{0}{b'}}{\varphi_2}{\tau}\right) && \text{\footnotesize (by global assump.)} \\
               =&\MonU{\bwp{0}{b'}}{\varphi_1\lor\varphi_2}{\tau} &&\text{\footnotesize (by Def.~\ref{def:classicMonitor})}
            \end{align*}    
        \end{compactitem}
        

        \item Case $\varphi' = \Box_I\varphi$: By Def.~\ref{def:quanFineMonitor}, Def.~\ref{def:classicMonitor} and the assumption, it holds that:
        \begin{align*}
            &\inf_{t\in[0,b]}\InsRobVio{\bwp{0}{t}}{\Box_I\varphi}{\tau} \\
            =& \inf_{t\in[0,b]}\left(\inf_{\tau'\in\tau+I}\InsRobVio{\bwp{0}{t}}{\varphi}{\tau'}\right) && \text{\footnotesize (by Def.~\ref{def:quanFineMonitor})}\\
            =& \inf_{\tau'\in\tau+I}\left(\inf_{t\in[0,b]}\InsRobVio{\bwp{0}{t}}{\varphi}{\tau'}\right)&&\text{\footnotesize (by def. of inf)} \\
            =& \inf_{\tau'\in\tau+I}\left(\MonU{\bwp{0}{b}}{\varphi}{\tau'}\right) &&\text{\footnotesize (by assump.)}\\
            =& \MonU{\bwp{0}{b}}{\Box_I\varphi}{\tau} && \text{\footnotesize (by Def.~\ref{def:classicMonitor})}
        \end{align*}

    \item Case $\varphi' = \Diamond_I\varphi$: We prove Thm.~\ref{theo:reconstruct} holds for this case by induction on the length $b$ of the partial signal $\bwp{0}{b}$.
    \begin{compactitem}
        \item First, if $b = \tau$, it holds that:
        \begin{align*}
            &\inf_{t\in[0,b]}\InsRobVio{\bwp{0}{t}}{\Diamond_I\varphi}{\tau} \\
            =& \InsRobVio{\bwp{0}{\tau}}{\Diamond_I\varphi}{\tau} && \text{\footnotesize (by Def.~\ref{def:quanFineMonitor})}\\
            =& \max\left(\inf_{\tau'\in\tau+I}\MonU{\bwp{0}{\tau}}{\varphi}{\tau'}, \MonU{\bwp{0}{\tau}}{\Diamond_I\varphi}{\tau}\right) &&\text{\footnotesize (by Def.~\ref{def:quanFineMonitor})} \\
            =& \Rmax = \MonU{\bwp{0}{b}}{\Diamond_I\varphi}{\tau} &&\text{\footnotesize (by Def.~\ref{def:classicMonitor})}
        \end{align*}
        \item Then, we make a local assumption that, given an arbitrary $b$, it holds that $\inf_{t\in[0,b]}\InsRobVio{\bwp{0}{t}}{\Diamond_I\varphi}{\tau} = \MonU{\bwp{0}{b}}{\Diamond_I\varphi}{\tau}$. We prove that, for $b'$ which is the next sampling point to $b$, it holds that,
        {\mathcompact
        \begin{align*}
            &\inf_{t\in[0,b']}\InsRobVio{\bwp{0}{t}}{\Diamond_I\varphi}{\tau} \\
            =& \min\left(\MonU{\bwp{0}{b}}{\Diamond_I\varphi}{\tau}, \InsRobVio{\bwp{0}{b'}}{\Diamond_I\varphi}{\tau}\right) \tag{\footnotesize by local assump.}\\
            =&\min\left(
            \begin{array}{lccccl}
               \max\Big( &\MonU{\bwp{0}{b}}{\varphi}{\tau},  & \ldots & \ldots & \MonU{\bwp{0}{b}}{\varphi}{\sup(\tau+I)} &\Big)\\
                \max\Big( &\InsRobVio{\bwp{0}{b'}}{\varphi}{\tau}, & \MonU{\bwp{0}{b'}}{\varphi}{\tau+\stepLength}, &\ldots & \MonU{\bwp{0}{b'}}{\varphi}{\sup(\tau+I)} &\Big) \\
                \max\Big(&\MonU{\bwp{0}{b'}}{\varphi}{\tau}, & \InsRobVio{\bwp{0}{b'}}{\varphi}{\tau+\stepLength}, & & \vdots&\\
                &\vdots& & \ddots & \vdots & \\
                \max\Big(&\MonU{\bwp{0}{b'}}{\varphi}{\tau}, & \ldots &\ldots  & \InsRobVio{\bwp{0}{b'}}{\varphi}{\sup(\tau+I)}&\Big)
            \end{array}
            \right)  \tag{\footnotesize by Def.~\ref{def:classicMonitor} and Def.~\ref{def:quanFineMonitor}}\\
            =& \max\left(
            \begin{array}{l}
                 \min\left(\MonU{\bwp{0}{b}}{\varphi}{\tau}, \InsRobVio{\bwp{0}{b'}}{\varphi}{\tau}\right)  \\
                \vdots \\
                \min\left(\MonU{\bwp{0}{b}}{\varphi}{\sup(\tau+I)}, \InsRobVio{\bwp{0}{b'}}{\varphi}{\sup(\tau+I)}\right)
            \end{array}
            \right) \tag{\footnotesize by def. of min, max}\\
            =& \sup_{t\in\tau+I}\MonU{\bwp{0}{b'}}{\varphi}{t} \tag{\footnotesize by global assump.}\\
            =& \MonU{\bwp{0}{b'}}{\Diamond_I\varphi}{\tau} \tag{\footnotesize by Def.~\ref{def:classicMonitor}}
        \end{align*}}
    \end{compactitem}
    \item Case $\varphi' = \varphi_1\UntilOp{I}\varphi_2$: We prove Thm.~\ref{theo:reconstruct} holds for this case by induction on the length $b$ of the partial signal $\bwp{0}{b}$.
    \begin{compactitem}
        \item First, if $b = \tau$, it holds that:
        \begin{align*}
            &\inf_{t\in[0,b]}\InsRobVio{\bwp{0}{t}}{\varphi_1\UntilOp{I}\varphi_2}{\tau} \\ 
            =& \InsRobVio{\bwp{0}{\tau}}{\varphi_1\UntilOp{I}\varphi_2}{\tau} \tag{\footnotesize by Def.~\ref{def:quanFineMonitor}}\\
            =& \inf_{\tau'\in\tau+I}\left(
    \max\left(
    \begin{array}{l}
         \min\left(
         \begin{array}{l}
              \displaystyle\inf_{t'\in[\tau,\tau')}\InsRobVio{\bwp{0}{\tau}}{\varphi_1}{t'}  \\
               \InsRobVio{\bwp{0}{\tau}}{\varphi_2}{\tau'}
         \end{array}
         \right)  \\
          \MonU{\bwp{0}{\tau}}{\varphi_1\UntilOp{I}\varphi_2}{\tau}
    \end{array}
    \right)
    \right) \tag{\footnotesize by Def.~\ref{def:quanFineMonitor}} \\
    =&\max\left(
    \begin{array}{l}
         \InsRobVio{\bwp{0}{\tau}}{\varphi_2}{\tau}  \\
        \MonU{\bwp{0}{\tau}}{\varphi_1\UntilOp{I}\varphi_2}{\tau}
    \end{array}
    \right)   \tag{\footnotesize by Def.~\ref{def:quanFineMonitor}}\\
    =& \MonU{\bwp{0}{b}}{\varphi_1\UntilOp{I}\varphi_2}{\tau} \tag{\footnotesize by local assump. and Def.~\ref{def:classicMonitor}}
        \end{align*}
        \item Then, we make a local assumption that, given an arbitrary $b$, it holds that $\inf_{t\in[0,b]}\InsRobVio{\bwp{0}{t}}{\varphi_1\UntilOp{I}\varphi_2}{\tau} = \MonU{\bwp{0}{b}}{\varphi_1\UntilOp{I}\varphi_2}{\tau}$. We prove that, for $b'$ which is the next sampling point to $b$, it holds that,
        {\mathcompact
        \begin{align*}
            &\inf_{t\in[0,b']}\InsRobVio{\bwp{0}{t}}{\varphi_1\UntilOp{I}\varphi_2}{\tau} \\
            =&\min\left(\MonU{\bwp{0}{b}}{\varphi_1\UntilOp{I}\varphi_2}{\tau}, \InsRobVio{\bwp{0}{b'}}{\varphi_1\UntilOp{I}\varphi_2}{\tau}\right) \tag{\footnotesize by local assump.} \\
            =& \min\left(
            \begin{array}{lcccl}
                \max\Bigg( & \min\left(
                \begin{array}{l}
                     \MonU{\bwp{0}{b}}{\varphi_2}{\tau}  \\
                    \displaystyle\inf_{t'\in[\tau, \tau]}\MonU{\bwp{0}{b}}{\varphi_1}{t'}
                \end{array}
                \right)  &\ldots & \min\left(
                \begin{array}{l}
                     \MonU{\bwp{0}{b}}{\varphi_2}{\tau+I}  \\
                    \displaystyle\inf_{t'\in\tau+I}\MonU{\bwp{0}{b}}{\varphi_1}{t'}
                \end{array}
                \right) & \Bigg) \\ 
                \max\Bigg( & 
                \min\left(
                \begin{array}{l}
                     \InsRobVio{\bwp{0}{b'}}{\varphi_2}{\tau}  \\
                    \displaystyle\inf_{t'\in[\tau,\tau]}\InsRobVio{\bwp{0}{b'}}{\varphi_1}{t'}
                \end{array}
                \right)  & \ldots& 
                \min\left(
                \begin{array}{l}
                     \MonU{\bwp{0}{b'}}{\varphi_2}{\tau+I}  \\
                    \displaystyle\inf_{t'\in\tau+I}\MonU{\bwp{0}{b'}}{\varphi_1}{t'}
                \end{array}
                \right) & \Bigg) \\
                 &\vdots &\ddots & \vdots&  \\ 
                \max\Bigg( & 
                \min\left(
                \begin{array}{l}
                     \MonU{\bwp{0}{b'}}{\varphi_2}{\tau}  \\
                    \displaystyle\inf_{t'\in[\tau,\tau]}\MonU{\bwp{0}{b'}}{\varphi_1}{t'}
                \end{array}
                \right)  &\ldots & 
                \min\left(
                \begin{array}{l}
                     \InsRobVio{\bwp{0}{b'}}{\varphi_2}{\tau+I}  \\
                    \displaystyle\inf_{t'\in\tau+I}\InsRobVio{\bwp{0}{b'}}{\varphi_1}{t'}
                \end{array}
                \right) &\Bigg)  
            \end{array}
            \right)  \tag{\footnotesize by Def.~\ref{def:classicMonitor} and Def.~\ref{def:quanFineMonitor}}\\
            =& \max\left(
            \begin{array}{l}
            \min\left(
                  \min\left(
                  \begin{array}{l}
                       \MonU{\bwp{0}{b}}{\varphi_2}{\tau}  \\
                       \displaystyle\inf_{t'\in[\tau, \tau]}\MonU{\bwp{0}{b}}{\varphi_1}{t'} 
                  \end{array}
                  \right), \min\left(
                    \begin{array}{l}
                         \InsRobVio{\bwp{0}{b'}}{\varphi_2}{\tau}  \\
                         \displaystyle\inf_{t'\in[\tau, \tau]}\InsRobVio{\bwp{0}{b'}}{\varphi_1}{t'} 
                    \end{array}
                  \right)
            \right) \\
                  \vdots \\
                  \min\left(
                  \min\left(
                  \begin{array}{l}
                       \MonU{\bwp{0}{b}}{\varphi_2}{\tau+I}  \\
                       \displaystyle\inf_{t'\in\tau+I}\MonU{\bwp{0}{b}}{\varphi_1}{t'} 
                  \end{array}
                  \right), \min\left(
                  \begin{array}{l}
                       \InsRobVio{\bwp{0}{b'}}{\varphi_2}{\tau+I}  \\
                       \displaystyle\inf_{t'\in\tau+I}\InsRobVio{\bwp{0}{b'}}{\varphi_1}{t'} 
                  \end{array}
                  \right)
                  \right)
            \end{array}
            \right) \tag{\footnotesize by def. of min, max}\\
            =& \sup_{t\in\tau +I}\min\left(\MonU{\bwp{0}{b'}}{\varphi_2}{t}, \inf_{t'\in[\tau, t)}\MonU{\bwp{0}{b'}}{\varphi_1}{t'}\right) \hspace*{-0.1cm}\tag{\footnotesize by global assump.}\\
            =& \MonU{\bwp{0}{b'}}{\varphi_1\UntilOp{I}\varphi_2}{\tau} \tag{by Def.~\ref{def:classicMonitor}}
        \end{align*}}
    \end{compactitem}
    \end{compactitem}

The proof for $\sup_{t\in[0,b]}\InsRobSat{\bwp{0}{t}}{\varphi'}{\tau} = \MonL{\bwp{0}{b}}{\varphi'}{\tau}$ follows a similar pattern, and so we skip it. \qed
\end{proof}

\end{document}